\documentclass[pra,%
 reprint,
 amsmath,amssymb,
 aps,
]{revtex4-2}

\usepackage{graphicx}
\usepackage{xcolor}

\usepackage{amsmath}
\usepackage{mathtools}

\usepackage{dcolumn}
\usepackage{bm}
\usepackage{float}
\usepackage{physics}

\begin{document}

\preprint{APS/123-QED}

\title{Non-Hermitian off-diagonal disordered optical lattices}

\author{E. T. Kokkinakis$^{1,2}$}

\author{I. Komis $^{1,2}$}

\author{K. G. Makris$^{1,2}$}%

\author{E. N. Economou$^{1,2}$}
\affiliation{$^1$
  Department of Physics, University of Crete, 70013 Heraklion, Greece
}%
\affiliation{$^2$
 Institute of Electronic Structure and Laser (IESL), FORTH, 71110 Heraklion, Greece
}%
\date{\today}

\begin{abstract}
Within the framework of non-Hermitian photonics, we investigate the spectral and dynamical properties of one- and two-dimensional non-Hermitian off-diagonal disordered optical lattices, where randomness is applied to the couplings rather than to the on-site potential terms. We analyze eigenvalue distributions and the localization properties of the eigenmodes, comparing them with those of the corresponding Hermitian lattices. Furthermore, we study their transport behavior under single-channel excitation and identify unconventional phenomena such as jumps between distant lattice regions in systems with a purely real spectrum, as well as complex spectrum-induced Anderson jumps, reported here for the first time in two dimensions. Our results establish a reference framework for non-Hermitian off-diagonal disorder and open new directions for future studies of localization phenomena.
\end{abstract}

\maketitle

\section{INTRODUCTION}
The concept of Anderson localization has attracted central attention in condensed-matter physics for decades \cite{Lee_Ramakrishnan_1985, Evers_Mirlin_2008}, fundamentally shaping our understanding of wave transport in disordered materials. Following Anderson’s seminal work in 1958 \cite{Anderson_1958}, early studies \cite{Abrahams_1979, Thouless_1972, MacKinnon_1981, Anderson_Thouless_Abrahams_Fisher_1980} established that one-dimensional (1D) and two-dimensional (2D) infinite lattices exhibit localized eigenmodes for arbitrarily small on-site uncorrelated disorder, whereas in three-dimensional (3D) systems the spectrum contains mobility edges separating regions of extended and localized states. In the latter case, an Anderson transition occurs at a critical disorder strength, beyond which all eigenmodes become localized.

Although most studies of Anderson localization have concentrated on on-site randomness, the role of off-diagonal disorder—random couplings with uniform on-site energies—was already explored by the late 1970s. In 1D and 2D random-hopping systems, all states are localized in the usual sense; however, the mid-band (zero-energy) state exhibits a divergent localization length and density of states, remaining non-metallic and sub-exponentially localized \cite{Theodorou_1976, Eggarter_1978, Soukoulis_1981, Fleishman_1977, Ziman_1982, Soukoulis_1982}. These features were later recognized as generic for bipartite lattices, disappearing when on-site disorder is introduced \cite{Stone_1981a, Stone_1981b, Evangelou_1986} or when next-nearest-neighbor terms are added \cite{Inui_1994}. In 3D, by contrast, the random-hopping Anderson model retains extended states around the band center for any hopping-disorder strength; as disorder grows, a finite mid-band window of extended states persists \cite{Economou_1977, Antoniou_1977, Biswas_2000}. Moreover, several studies showed that correlated off-diagonal disorder in 1D can induce localization–delocalization transitions, yielding extended states even in the presence of on-site random binary disorder \cite{Cheraghchi_2005, Zhang_2004}.

Although Anderson localization was originally formulated in the context of solid-state physics, its experimental observation has been more easily accessible in other areas of physics where many-body interactions and temperature-dependent effects are absent. These include cold-atoms and Bose-Einstein condensates \cite{damski_2003, billy_2008, roati_2008, chabe_2008, jendrzejewski_2012}, plasmonics \cite{cherpakova_2017, shi_2018, pandey_2017, zhu_2020, zhu_2020, duan_2019}, acoustics \cite{condat_1987, hu_2008}, and, most prominently, photonics \cite{wiersma_2013, schwartz_2007, lahini_2008, martin_2011, stutzer_2012, karbasi_2012}. In photonics, Anderson localization can be realized by introducing controlled randomness into the refractive-index distribution, for example in laser-written waveguide arrays. Notably, experiments on {off-diagonal} disorder have been also enabled in such photonic platforms by randomizing the spacings between adjacent waveguides \cite{martin_2011}.

Relatively recently, research on Anderson localization has extended to non-Hermitian systems, motivated by the first experimental observation of parity–time symmetry in optics \cite{Ruter2010, Makris2008, ElGanainy2007, Musslimani2008, Guo2009, Regensburger2012, Hodaei2014, Konotop2016, Feng2017, Ozdemir2019}, which firmly established non-Hermitian physics as a modern research frontier \cite{ElGanainy2018}. This development has spurred renewed interest in non-Hermitian disordered problems \cite{makris_2017, makris_2020, tzortzakakis_2020, huang_2020, tzortzakakis_2020_2, Sukhachov_2020, liu_2020, kawabata_2021, liu_2020_b, leventis_2022, acharya_2022, weidemann_2021, longhi_2023_adp, tzortzakakis_2021, liu_2021_3, liu_2021_2, luo_2021, luo_2021_2, luo_2022, he_2024, sun_2024, li_2024, wang_2024, Kokkinakis_2024, kokkinakis_2025, Komis2025Evolving, Kokkinakis2025IncoherentSkin, Komispra2025}, where the interplay of disorder and properly engineered gain/loss profiles can be investigated experimentally.
Among the variety of intricate phenomena revealed by recent research on non-Hermitian Anderson localization, we refer as representative examples to constant-intensity waves persisting under strong localization \cite{makris_2017, makris_2020}, scale-free localized states \cite{LiLeeGong2021, Yokomizo2021, Liu2021, Wang2022, FuZhang2023, LiSunYangLi2024, Yuce2025, Molignini2023, Liu2024, Yilmaz2024, Y_Zhang2025, Li2023, Guo2023, Kokkinakis2025NHImpurity}, and abrupt Anderson jumps \cite{tzortzakakis_2021, weidemann_2021, leventis_2022, longhi_2023_adp, kokkinakis_2025, Kokkinakis2025IncoherentSkin, Komis2025Evolving}.

However, despite the growing interest in non-Hermitian disordered physics and photonics, systematic studies of non-Hermitian systems with off-diagonal disorder remain limited, mainly focusing on quasiperiodic lattices with incommensurate potentials \cite{Liu_2018, Krishna_2020, Zhang_2025}. The purpose of this work is to address this gap by examining the spectral and dynamical properties of finite one- and two-dimensional optical lattices with random couplings. In particular, we analyze the distribution of eigenvalues in the complex plane and the localization properties of the associated states. Regarding the dynamics, we investigate transport under single-channel excitation and provide a direct comparison with the corresponding Hermitian systems. We believe that our work is relevant to the broader non-Hermitian disordered physics community, and particularly to experimental implementations in photonics, where non-Hermitian random couplings can be physically realized.

\section{ONE-DIMENSIONAL OFF-DIAGONAL DISORDERED LATTICES} 
\subsection{Hermitian and non-Hermitian models}
Our study begins with a 1D lattice consisting of $N$ evanescently coupled waveguides (indexed by $n \in \{ 1,2,...,N \}$) that have uniform on-site terms $\alpha_n\equiv 0$ and site-dependent nearest-neighbor couplings. Thus, the Hamiltonian of the system is described by
\begin{equation}
\label{off}
H = \sum_{n=1}^{N-1} \Big( t_{L,n}\,|n\rangle\langle n+1| + t_{R,n}\,|n+1\rangle\langle n| \Big) 
\end{equation}
where $t_{L,n}$ and $t_{R,n}$ denote the coupling amplitudes to the backward and forward directions, respectively. We consider that the waveguides form a linear chain with open boundary conditions (OBC) hold, i.e., $\ket{0}=\ket{N+1}\equiv0$. In what follows, we distinguish between three different cases depending on the specific values of the coupling coefficients $t_{L,n}$ and $t_{R,n}$.

In the \emph{Hermitian} off-diagonal disorder case, the couplings are identical in both directions, $t_{L,n}=t_{R,n}\equiv t_n$, where $t_n$ is a random positive real number drawn from a uniform distribution $t_n\in\big[1-\tfrac{W}{2},\,1+\tfrac{W}{2}\big]$. The parameter $W$ characterizes the disorder strength and, by convention, is restricted to the interval $W\in[0,2]$ to ensure $t_n \geq 0$. As shown in Appendix~A, in this case the eigenvalue problem $H\ket{u_j}=\epsilon_j \ket{u_j}$ yields a \emph{chiral} spectrum $\{\epsilon_j\}$, i.e., a real spectrum symmetric with respect to $\epsilon=0$. In particular, if $\epsilon$ is an eigenvalue of $H$ with eigenvector $\ket{\xi}$, then $-\epsilon$ is also an eigenvalue with eigenvector $\ket{\phi}$.  Moreover, the corresponding eigenstates $\ket{\xi}$ and $\ket{\phi}$ exhibit identical amplitude distributions across the lattice $\abs{\xi_n} \equiv \abs{\braket{n}{\xi}} = \abs{\braket{n}{\phi}} \equiv \abs{\phi_n}, \quad \forall n$.
\begin{figure}[t]
    \centering
    \includegraphics[width=1\columnwidth]{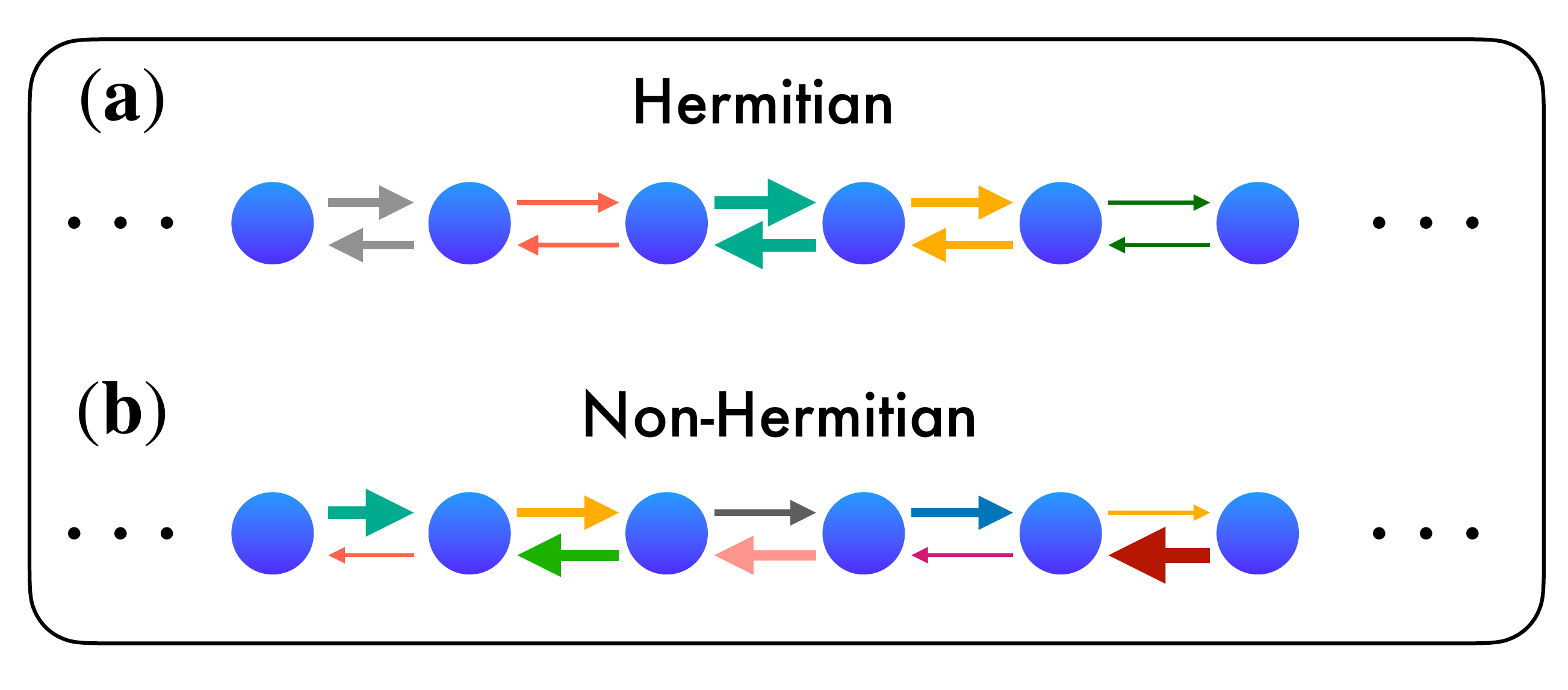}
    \caption{\textbf{Schematic representation} of the one-dimensional lattice models considered in this work.  (a) Random nearest-neighbor couplings are identical in the forward and backward directions, leading to a Hermitian Hamiltonian.  (b) Random nearest-neighbor couplings differ between forward and backward directions, giving rise to a non-Hermitian Hamiltonian. In both cases, variations in the color and width of the arrows indicate different coupling strengths, while in case (b) the couplings may be either real or complex.}
\end{figure}
Turning to the non-Hermitian cases, we first examine the situation where the couplings in the two directions remain positive and real but are drawn independently from two separate uniform distributions, namely $t_{L,n}\in\big[1-\tfrac{W}{2},\,1+\tfrac{W}{2}\big]$ and $t_{R,n}\in\big[1-\tfrac{W}{2},\,1+\tfrac{W}{2}\big]$ with $W\in [0,2]$. In this scenario the Hamiltonian $H$ remains real-valued, yet it is non-Hermitian due to the asymmetry of the couplings, i.e., $t_{L,n}\neq t_{R,n}$ in general. We will refer to this configuration as the \emph{real non-Hermitian} model. Remarkably, as discussed in Appendix~A, under OBC this Hamiltonian possesses an entirely real spectrum despite its non-Hermitian nature, while the spectrum also retains chiral symmetry. The right eigenvalue problem associated with this Hamiltonian is expressed as  ${H}\ket{u_{j}^{R}} = \epsilon_{j}\ket{u_{j}^{R}}$. Since $H$ is real, the relations ${H}^{\dagger}={H}^{T}$ and $\epsilon_j=\epsilon_{j}^{*}$ hold, so its left eigenvalue problem reads as ${H}^{T}\ket{u_{j}^{L}} = \epsilon_{j}\ket{u_{j}^{L}}$. The left and right eigenvectors are related through the biorthogonality condition
$\braket{u_{k}^{L}}{u_{j}^{R}} = \delta_{kj}$.

Finally, we also consider the case where the couplings between forward and backward directions are both unequal and complex, defined as
\[
t_{L,n}=\frac{a_n+ib_n}{\sqrt{2}}, \qquad 
t_{R,n}=\frac{c_n-id_n}{\sqrt{2}},
\]
where $a_n$, $b_n$, $c_n$, and $d_n$ are independent random real variables drawn from the uniform distribution $\big[1-\tfrac{W}{2},\,1+\tfrac{W}{2}\big]$, with $W\in[0,2]$. This choice of distribution ensures that in the limit $W\to 0$ one recovers $t_{L,n}=t_{R,n}^{*}$, and therefore the Hamiltonian becomes Hermitian. The prefactors $1/\sqrt{2}$ are introduced to guarantee that the coupling magnitudes satisfy $|t_{L/R,n}|<2$, thereby maintaining consistency with the bounds used in the previous configurations. In this case the spectrum of the Hamiltonian is genuinely complex, $\epsilon_j \in \mathbb{C}$; nevertheless, the chiral symmetry persists in the complex plane, with eigenvalues appearing in symmetric pointwise pairs with respect to the origin, having equal amplitudes at each lattice site. We will refer to this configuration as the \emph{complex non-Hermitian} model.

In the following subsections, we analyze and compare the spectral and transport properties of the finite-length ($N$) 1D off-diagonal disordered lattices introduced above. While the primary focus of this work is on non-Hermitian disordered lattices, we also include the Hermitian model for reference and comparison. A schematic representation highlighting the difference of the Hermitian and the non-Hermitian lattices is shown in Fig. 1. 
\subsection{Spectral properties}
In this subsection, we examine the spectral properties of the finite-sized lattices with non-Hermitian off-diagonal disorder, focusing on the eigenvalue's distribution along the complex plane (or real axis) as well as the degree of localization of the associated eigenmodes. 
\begin{figure*}[t]
    \centering
    \includegraphics[width=0.68\textwidth]{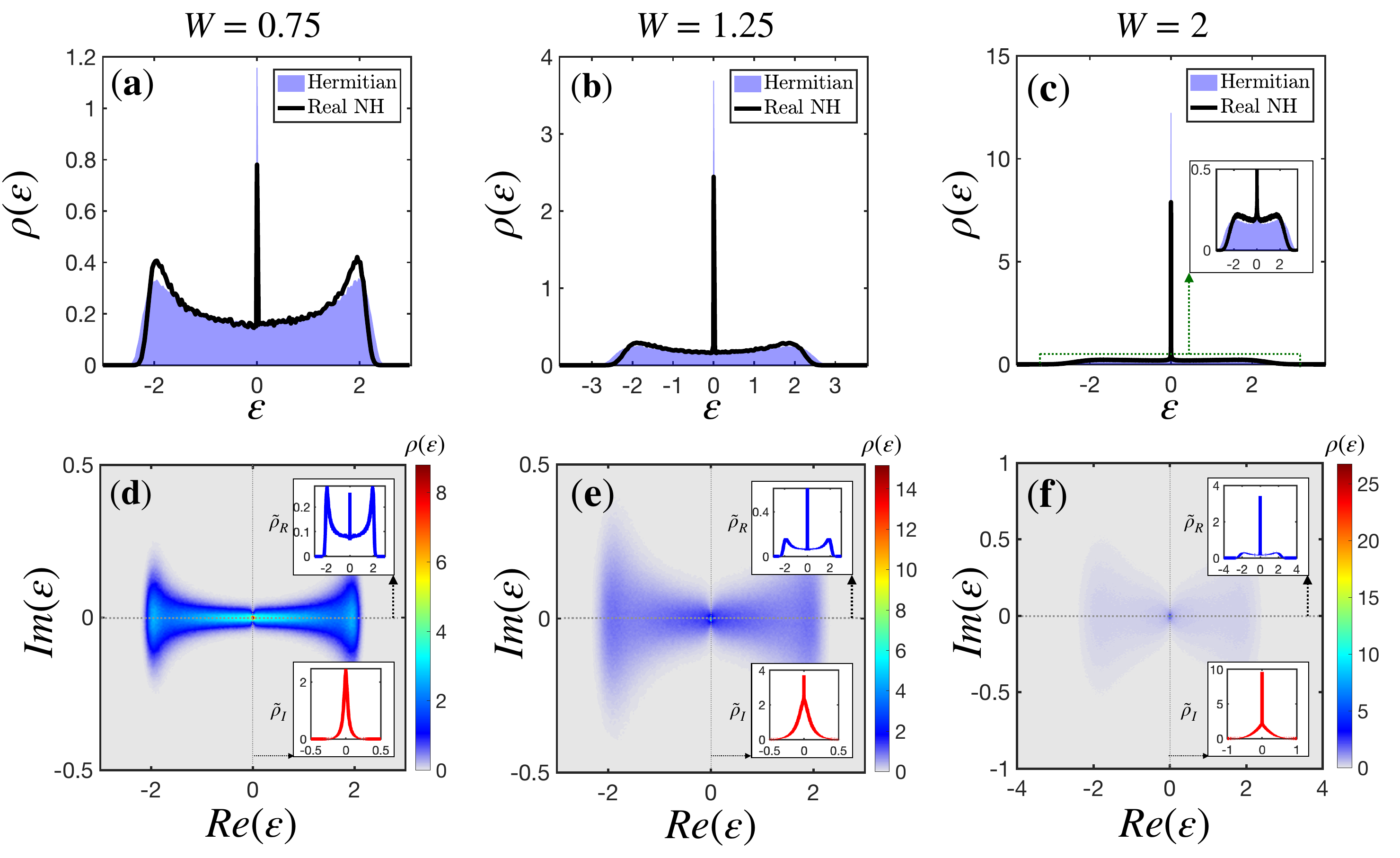}
    \caption{\textbf{Density of states for 1D lattices} ($N=500$ sites): (a)–(c) Comparison of the DOS of real eigenvalues, $\rho(\epsilon)$, between the \textit{Hermitian} (blue-shaded surface) and \textit{real non-Hermitian} (black line) cases for (a) $W=0.75$, (b) $W=1.25$, and (c) $W=2$. (d)–(f) DOS in the complex plane (colormap) for the \textit{complex non-Hermitian} case. \textit{Insets} in bottom row panels show the real-projected DOS $\tilde{\rho}_R$ (blue lines) and the imaginary-projected DOS $\tilde{\rho}_I$ (red lines) for (d) $W=0.75$, (e) $W=1.25$, and (f) $W=2$.}
    \label{fig_2}
\end{figure*}
Regarding the eigenvalue distribution of a disordered Hamiltonian, the most useful metric is its \emph{density of states} (DOS), $\rho(\epsilon)$, which for systems with real spectra ($\epsilon \in \mathbb{R}$ is defined as 
\begin{equation}
    \rho(\epsilon)\propto \frac{dN}{d\epsilon}
\end{equation}
where $dN$ is the number of eigenmodes with eigenvalue in the interval $\beta(\epsilon)=[\epsilon-d\epsilon/2,\epsilon+d\epsilon/2]$, \emph{averaged} over numerous disorder realizations. We conventionally choose the normalization of $\rho(\epsilon)$ to be such that $\int_{-\infty}^{+\infty} \rho(\epsilon) d \epsilon =1$. 

For systems with complex spectra ($\epsilon \in \mathbb{C}$), the DOS is generalized to the complex plane and is defined as 
\begin{equation}
    \rho(\epsilon)\propto \frac{dN}{dA},
\end{equation}
where $dN$ is the number of eigenmodes with eigenvalue lying inside the area element $dA=d\Re(\epsilon)\, d\Im(\epsilon)$ centered at $\epsilon=\Re(\epsilon)+i\Im(\epsilon)\equiv \epsilon_R+i\epsilon_I$, again \emph{averaged} over many disorder realizations. The normalization is chosen such that $\iint_{\mathbb{C}} \rho(\epsilon)\, dA = 1.$

In the complex-spectrum case, it is often useful to consider the \emph{projected} DOS for the real ($\tilde{\rho}_R$) and imaginary parts ($\tilde{\rho}_I$) of the eigenvalues. We define
\begin{equation}
    \tilde{\rho}_{R/I}(\epsilon_{R/I}) \propto \frac{dN}{d\epsilon_{R/I}},
\end{equation}
where $dN$ measures the number of eigenmodes with $\epsilon_{R/I}$ in $[\epsilon_{R/I}-d\epsilon_{R/I}/2,\;\epsilon_{R/I}+d\epsilon_{R/I}/2]$ irrespective of the complementary coordinate. Equivalently, $\tilde{\rho}_{R/I}$ are related to the complex plane DOS through the relation,
\begin{equation}
    \tilde{\rho}_{R/I}(\epsilon_{R/I}) \;=\; \int_{-\infty}^{+\infty} \rho\!\big(\epsilon\big)\; d\epsilon_{I/R},
\end{equation}
with the normalization $\int_{-\infty}^{+\infty} \tilde{\rho}_{R/I}(\epsilon_{R/I})\, d\epsilon_{R/I} \;=\; 1$.

To begin with, in the top row of Fig.~2 we compare the DOS of real eigenvalues between the Hermitian and the real non-Hermitian lattice models for different disorder strengths in lattices of size $N=500$. For both cases, the DOS at weak disorder $W=0.75$ [Fig.~1(a)] exhibits two mild peaks near $|\epsilon|=2$, corresponding to the edges of the unperturbed ($W=0$) band and reminiscent of the associated Van Hove singularities.  Most importantly, in both models a pronounced peak emerges near $\epsilon=0$. This central feature grows rapidly as the disorder strength increases to $W=1.25$ [Fig.~1(b)] and eventually dominates the spectrum at strong disorder $W=2$ [Fig.~1(c)]. The overall differences between the DOS of the Hermitian and real non-Hermitian lattices are relatively minor, primarily reflected in the height of the central peak and the precise form of the tails, without any substantial qualitative difference, an outcome that might initially appear unexpected. However, as discussed in Appendix~A, the Hamiltonian $H$ of a real non-Hermitian off-diagonal disordered lattice with couplings $t_{L,n}$ and $t_{R,n}$ is related by a similarity transformation, and hence is isospectral, to a \emph{Hermitian} off-diagonal disordered matrix $S$ with reciprocal couplings $\tau_{n}=\sqrt{t_{L,n}t_{R,n}}$. Consequently, the DOS of the real non-Hermitian model coincides with that of a Hermitian system with a modified, \textit{non-uniform} distribution of couplings. The qualitative similarity of the DOS in the two models is therefore a natural result, whereas, as will be shown in the next subsection, their transport properties exhibit far more significant differences.

\begin{figure*}[t]
    \centering
    \includegraphics[width=0.68\textwidth]{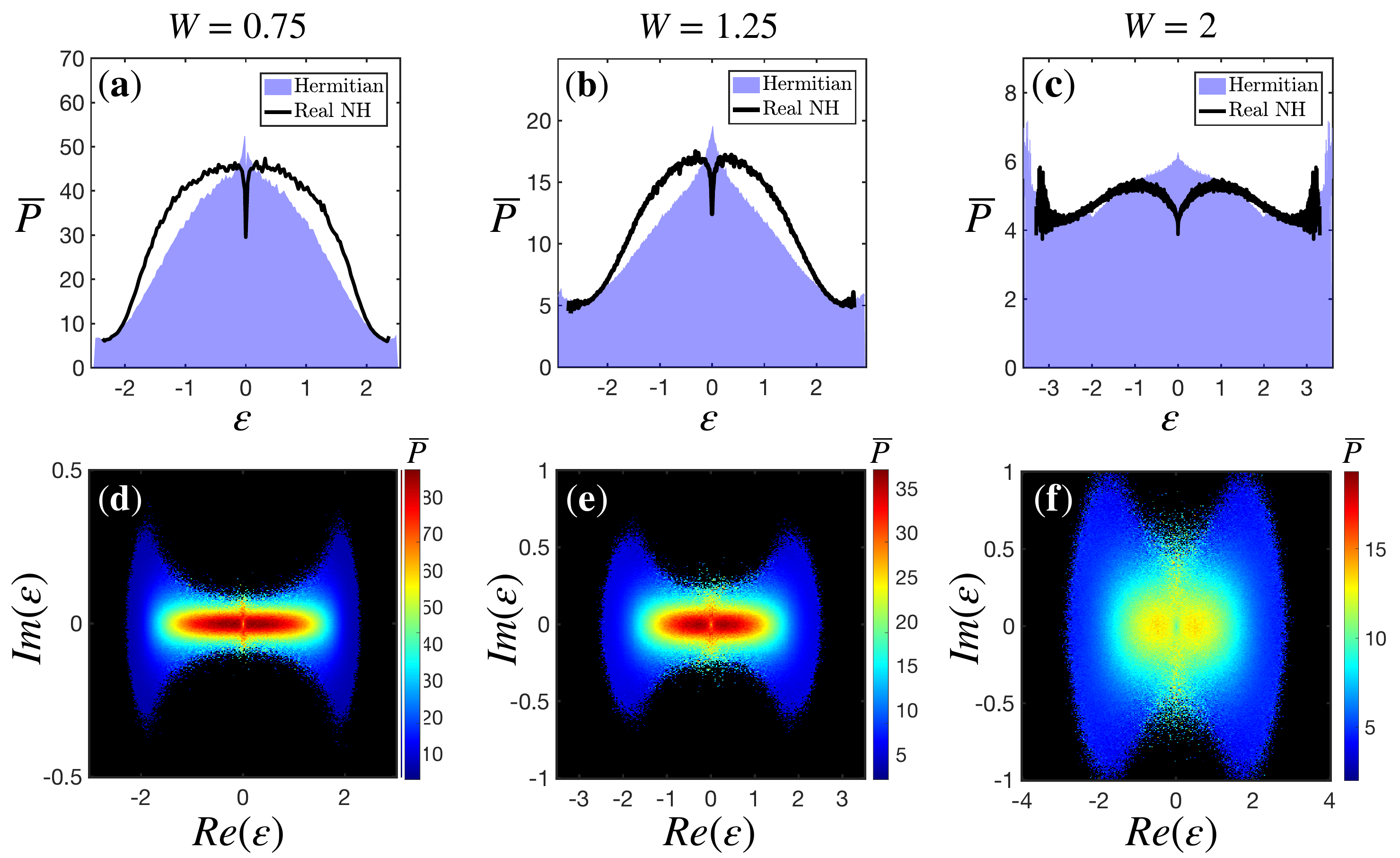}
    \caption{\textbf{Averaged participation ratios for 1D lattices}($N=500$ sites): (a)–(c) Comparison of $\overline{P}(\epsilon)$ for eigenmodes in the \textit{Hermitian} (blue-shaded surface) and \textit{real non-Hermitian} (black line) cases for (a) $W=0.75$, (b) $W=1.25$, and (c) $W=2$. (d)–(f) $\overline{P}(\epsilon)$ in the complex plane (colormap) for the \textit{complex non-Hermitian} case with (d) $W=0.75$, (e) $W=1.25$, and (f) $W=2$.}
    \label{fig_1}
\end{figure*}
The DOS acquires a more complicated form in the complex non-Hermitian model, shown in the bottom row of Fig.~2. At weak disorder $W=0.75$ [Fig.~2(d)], the complex-plane DOS attains its maximum near the origin $\epsilon=0$ and displays a two-lobed structure elongated along the real axis, with enhanced weight close to the unperturbed band edges $|\epsilon_R|\simeq 2$, as confirmed by the real-axis projected DOS $\tilde{\rho}_R$. By contrast, the imaginary-part distribution is much narrower; the imaginary-axis projected DOS $\tilde{\rho}_I$ is sharply peaked at $\epsilon_I=0$ and is nearly zero for $|\epsilon_I|\gtrsim 0.2$. Increasing the disorder to $W=1.25$ [Fig.~2(e)] further amplifies the central peak relative to the edge peaks, while $\tilde{\rho}_I$ broadens, extending up to $|\epsilon_I|\sim 0.4$. For strong disorder $W=2$ [Fig.~2(f)], the DOS becomes even more concentrated around the origin yet retains its double-lobe shape along the real axis; simultaneously, the imaginary-part distribution broadens further.

Regarding the degree of localization of the eigenmodes in disordered lattices, a widely used metric for finite-sized systems is the \emph{participation ratio} $P_{j}$ of the right eigenmodes $\ket{u^{R}_{j}}$ (or simply $\ket{u_{j}}$ in the Hermitian case), defined as
\begin{equation}
    \label{pr}
    P_{j} \equiv 
    \frac{\left( \sum_{n=1}^{N} \big| u^{R}_{j,n} \big|^2 \right)^{2}}
         {\sum_{n=1}^{N} \big| u^{R}_{j,n} \big|^{4}} ,
\end{equation}
where $u^{R}_{j,n}\equiv \braket{n}{u^{R}_{j}}$. This quantity ranges from $P_j=1$ for a state localized at a single site, i.e., $u^{R}_{j,n}=\delta_{n,n_0}$, to $P_j=N$ for a uniformly extended state across the lattice, i.e., $u^{R}_{j,n}=1/\sqrt{N}$. In principle, the scaling behavior of $P_j$ with system size $N$ provides information about localization in the thermodynamic limit $N\to \infty$. Here, however, we restrict our analysis to finite-sized lattices, which are relevant for implementations in optical platforms. To study how localization depends on the eigenvalue region, we define the averaged participation ratio $\overline{P}(\epsilon)$ as the average value of $P_j$ over all eigenmodes whose eigenvalues lie in a small region $\beta(\epsilon)$ around $\epsilon$, over numerous disorder realizations. For systems with purely real spectra, we take $\beta(\epsilon)=[\epsilon-d\epsilon/2,\,\epsilon+d\epsilon/2]$; for systems with complex spectra, we take a rectangular area $\beta(\epsilon)$ in the complex plane centered at $\epsilon=\Re(\epsilon)+i\,\Im(\epsilon)$. 
Denoting by $dN$ the total number of eigenvalues contained in $\beta(\epsilon)$ across all realizations, we define
\begin{equation}
    \overline{P}(\epsilon)
    = \frac{1}{dN}\!
      \sum_{\epsilon_{j}\in \beta(\epsilon)} P_j \, .
\end{equation}

For the Hermitian and real non-Hermitian cases, we compare the behavior of $\overline{P}(\epsilon)$ for lattices of $N=500$ in the top row of Fig.~3 for different disorder strengths $W$. In both lattices, $\overline{P}(\epsilon)$ decreases systematically across the spectrum as disorder increases, reflecting stronger localization: from typical values of $\sim 10\!-\!50$ at $W=0.75$ [Fig. 3(a)], to $\sim 5\!-\!20$ at $W=1.25$ [Fig. 3(b)], and down to $\sim 5$ at $W=2$ [Fig. 3(c)]. A notable difference, however, arises near $\epsilon=0$. In the Hermitian case, $\overline{P}(\epsilon)$ grows towards the band center, reaching its maximum at $\epsilon=0$. By contrast, in the real non-Hermitian case this monotonic trend is interrupted by a dip very close to the origin. In other words, for such lattices, for a fixed disorder strength, eigenmodes in the immediate vicinity of $\epsilon=0$ are on average more localized than those in adjacent spectral regions.

For the complex non-Hermitian lattice, on average the degree of localization is consistently weaker, with $\overline{P}(\epsilon)$ ranging from $\sim 10\!-\!85$ for $W=0.75$ [Fig.~3(d)] to $\sim 5\!-\!35$ for $W=1.25$ [Fig.~3(e)], and to $\sim 2\!-\!20$ for $W=2$ [Fig.~3(f)]. Remarkably, the dip of $\overline{P}(\epsilon)$ near $\epsilon=0$ persists for this model as well. This  feature, also observed in two-dimensional lattices discussed in the following section, appears to be a robust characteristic of non-Hermitian off-diagonal disordered lattices.

\subsection{Propagation dynamics}

Having examined the spectral properties of the one-dimensional off-diagonal disordered lattices, we now turn our attention to their transport features. The coupled-mode equation governing paraxial evolution in the lattice of Eq.~\eqref{off} reads,
\begin{equation}
    i\,\frac{d \psi_n}{dz}
    + t_{R,n}\,\psi_{n+1}
    + t_{L,n-1}\,\psi_{n-1}
    = 0
\end{equation}
or, in compact form   $i\,\partial_z\,\ket{\psi(z)} + H\,\ket{\psi(z)} = 0,
$
where $\psi_n(z)$ is the slowly varying envelope of the electric field at site $n$. For brevity we refer to $\ket{\psi(z)}=\sum_{n=1}^{N}\psi_n(z)\ket{n}$ as the wavefunction.  In the general non-Hermitian case, we expand the field in the right-eigenmode basis of $H$ as
\begin{equation}
    \label{exp}
    \psi_n(z)=\sum_{j=1}^{N} c_{j,0}\, e^{\,i\epsilon_{j}z}\, u^{R}_{j,n},
\end{equation}
with coefficients $c_{j,0}=\braket{u_{j}^{L}}{\psi(0)}$. In the Hermitian limit the decomposition \eqref{exp} holds with $\ket{u_j^{L}}=\ket{u_j^{R}}=\ket{u_j}$. By contrast, for a non-Hermitian system the optical power $\mathcal{P}(z)\equiv\bra{\psi}\ket{\psi}=\sum_{n=1}^{N}|\psi_n(z)|^2$ is not conserved. Thus, a normalized amplitude for the field's envelope can be introduced at every propagation distance $z$, namely
$\phi_{n}\equiv \psi_{n}/\sqrt{\mathcal{P}(z)}$. 

Some useful physical quantities concerning the transport properties of the system are, the mean position and its uncertainty, defined through the following relations:
\begin{equation}
	\langle n \rangle_{z} \equiv \sum_{n=1}^{N} n \left| \phi_{n}(z) \right|^2
\end{equation}

\begin{equation}
	 \Delta n _{z} \equiv \sqrt{\langle n^2 \rangle_{z} - \langle n \rangle_{z}^2}
\end{equation}
where $\langle n^2 \rangle_{z} \equiv \sum_{n=1}^{N} n^{2} \left| \phi_{n}(z) \right|^2$.

As is evident from Eq. \eqref{exp}, when the spectrum is real ($\epsilon_j\in\mathbb{R}$) the propagation dynamics arise from interference among right eigenmodes with weights set by the projections $c_{j,0}$, which depend solely on the initial condition and on the particular disorder realization. In practice, the evolution is dominated by modes $\ket{u_{j}^{R}}$ with large $|c_{j,0}|$, whereas eigenstates with $|c_{j,0}|$ that are negligible relative to the rest contribute only weakly to Eq. \eqref{exp}. A convenient indicator of the spatial location of each right eigenstate $\ket{u_{j}^{R}}$ along the lattice is its center of mass,
\begin{equation}
    \langle m \rangle_{j} \equiv \frac{\sum_{n=1}^{N} n\,\abs{u_{j,n}^{R}}^2}{\sum_{n=1}^{N} \abs{u_{j,n}^{R}}^2}.
\end{equation}
that will be used in what follows.
\begin{figure}[t]
    \centering
    \includegraphics[width=0.9\columnwidth]{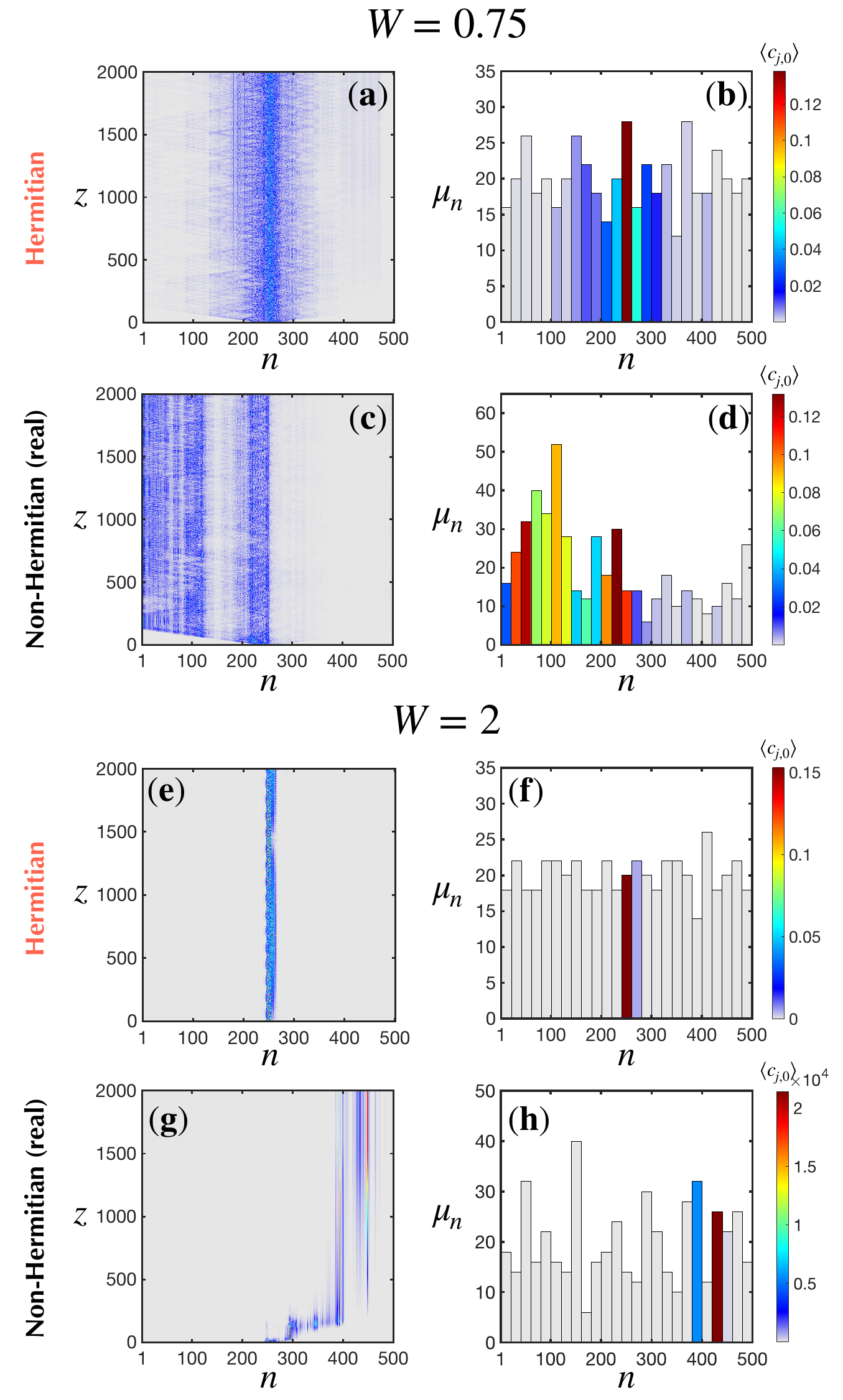}
    \caption{\textbf{Propagation dynamics for 1D lattices} ($N=500$ sites). Left panels: Evolution of the normalized wavefunction, initially localized at site $n=250$, for weak/strong disorder realizations ($W=0.75$ / $W=2$) in the (a)/(e) Hermitian and (c)/(g) real non-Hermitian cases. Right panels: For each lattice site $n$, the bar height gives the number $\mu_n$ of right eigenstates whose center of mass lies within the corresponding spatial bar. The color encodes the averaged projection coefficients $\langle |c_{j,0}|\rangle$ of the eigenmodes contained in each bin, for weak/strong disorder realizations ($W=0.75$ / $W=2$) in the (b)/(f) Hermitian and (d)/(h) real non-Hermitian cases.}

    \label{fig_1}
\end{figure}
We begin by examining the propagation dynamics for single realizations of Hermitian and real non-Hermitian off-diagonal disorder on a lattice with $N=500$ sites. Figure~4(a) illustrates the Hermitian case for a relatively weak disorder strength $W=0.75$. For a single-site initial excitation, $\psi_n(z=0)=\delta_{n,250}$, the wavefunction expands over a finite region of the lattice around $n=250$, yet it does not spread across the entire system, in agreement with the established behavior of Anderson localization. As expected, for a stronger disorder strength, $W=2$, the spatial extent of the wavefunction becomes even more confined, indicating a higher degree of localization [Fig. 4(c)].

In contrast, the dynamics are considerably more complex in the real non-Hermitian random–coupling case. As shown for a representative single realization in Fig.~4(c), part of the wavefunction remains near its initial excitation site, while another part is directed toward the left side of the lattice $(n=1)$, leading to a larger uncertainty $\Delta n(z)$ compared to the Hermitian case. The behavior becomes even more pronounced in the strong–disorder regime $(W=2)$. As illustrated in Fig.~4(g), the wavefunction is entirely displaced from its initial position and undergoes an abrupt jump. This jump is fundamentally different from previously reported \emph{non-Hermitian jumps} (also termed \emph{Anderson jumps}) in systems with complex spectra \cite{leventis_2022}, since, for the lattice under discussion, the spectrum of the Hamiltonian is entirely real. In this case, the wavefunction localizes around a different region of the lattice without exhibiting diffusion.

This pronounced qualitative difference between the reciprocal and non-reciprocal random–coupling models can be understood as follows: in the latter case, for certain disorder realizations, the presence of a preferential strong–coupling path can drive the wavefunction to delocalize toward a specific region of the lattice. Qualitatively similar features in the propagation dynamics have also been reported in on-site disordered Hatano–Nelson chains \cite{Kokkinakis_2024}. To analyze this difference more systematically, the right column of Fig.~4 presents, for each single disorder realization, the distribution $\mu_n$ of eigenstate centers of mass $\langle m \rangle_{j}$ along the lattice, grouped into a finite number of bins. The color of each bar represents the mean value of the projection coefficients $\langle |c_{j,0}| \rangle$ associated with the eigenmodes whose centers of mass fall within the corresponding bin.  

For the Hermitian cases, both at $W=0.75$ [Fig.~4(b)] and $W=2$ [Fig.~4(f)], the eigenmodes are distributed uniformly across the lattice, with the largest projection coefficients corresponding to modes located near the initial excitation site. In the low–disorder regime, the broader spreading of the wavefunction is reflected by nonzero projection coefficients over a finite region around the center of the lattice. By contrast, in the non-Hermitian cases the eigenmodes distribution exhibits higher asymmetry. At $W=0.75$ [Fig.~4(d)], non-negligible projection coefficients appear predominantly for modes located at $n<300$, with maximum values around $n=50$ and $n=230$. Even more strikingly, in the strong–disorder regime $W=2$ [Fig.~4(h)], the projection coefficients reach values several orders of magnitude larger in the regions near $n=380$ and $n=430$ compared to the rest of the lattice, in direct correspondence with the observed propagation dynamics. In both non-Hermitian examples, the values of the projection coefficients $|c_{j,0}|$ clearly indicate the asymptotic \emph{localization centers} of the wavefunction.
\begin{figure}[t]
    \centering
    \includegraphics[width=1\columnwidth]{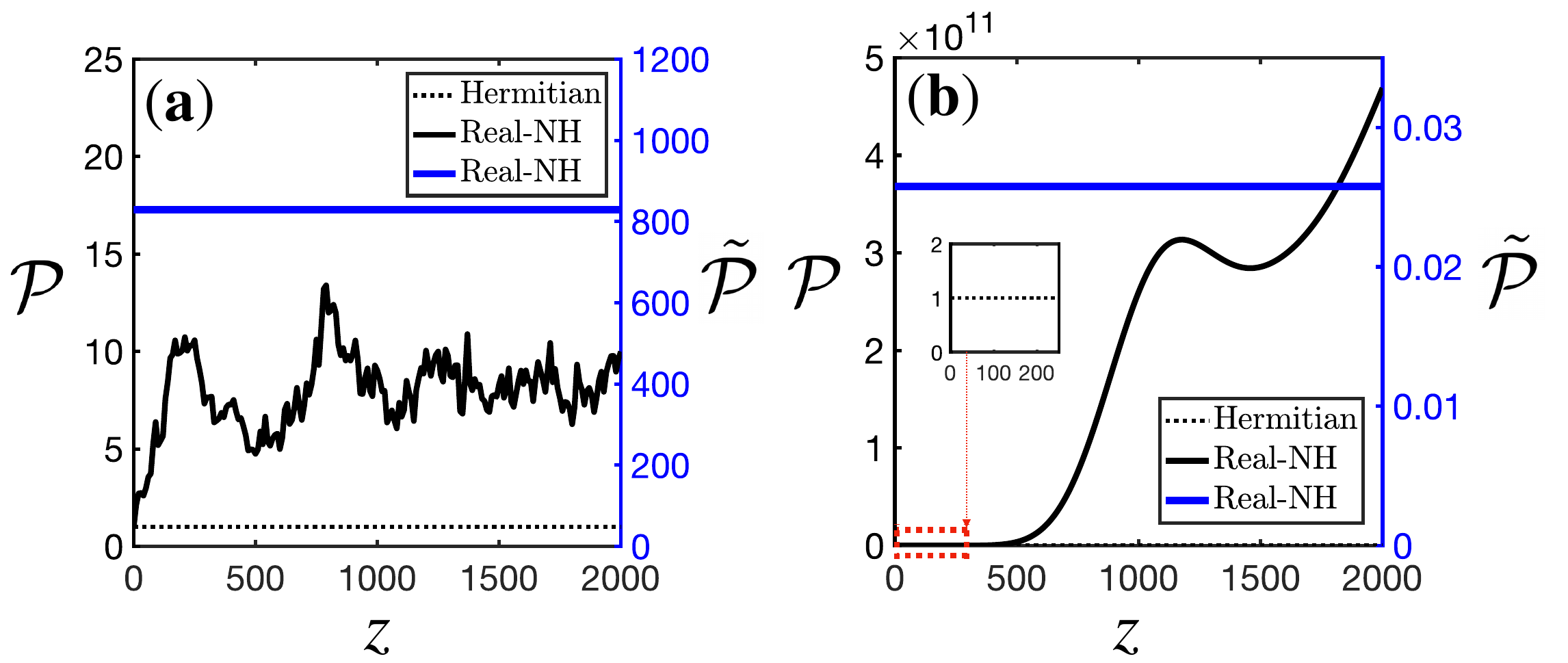}
    \caption{\textbf{Evolution of optical power:} (a) Comparison of the optical power $\mathcal{P}(z)$ for a single realization of weak disorder in the \textit{Hermitian} case (black dotted line) and the \textit{real non-Hermitian} case (black solid line). The evolution of the \textit{pseudopower} $\tilde{\mathcal{P}}$ [defined in Eq. (13)] for the latter case is also shown (blue solid line), shown to the right axis. (b) Same as in (a), but for single realizations of strong disorder ($W=2$). The Hermitian / non-Hermitian realizations used in panels (a) and (b) are the same as those in Figs.~4(a)/(c) and 4(e)/(g), respectively.}

    \label{fig_1}
\end{figure}
\begin{figure}[t]
    \centering
    \includegraphics[width=1\columnwidth]{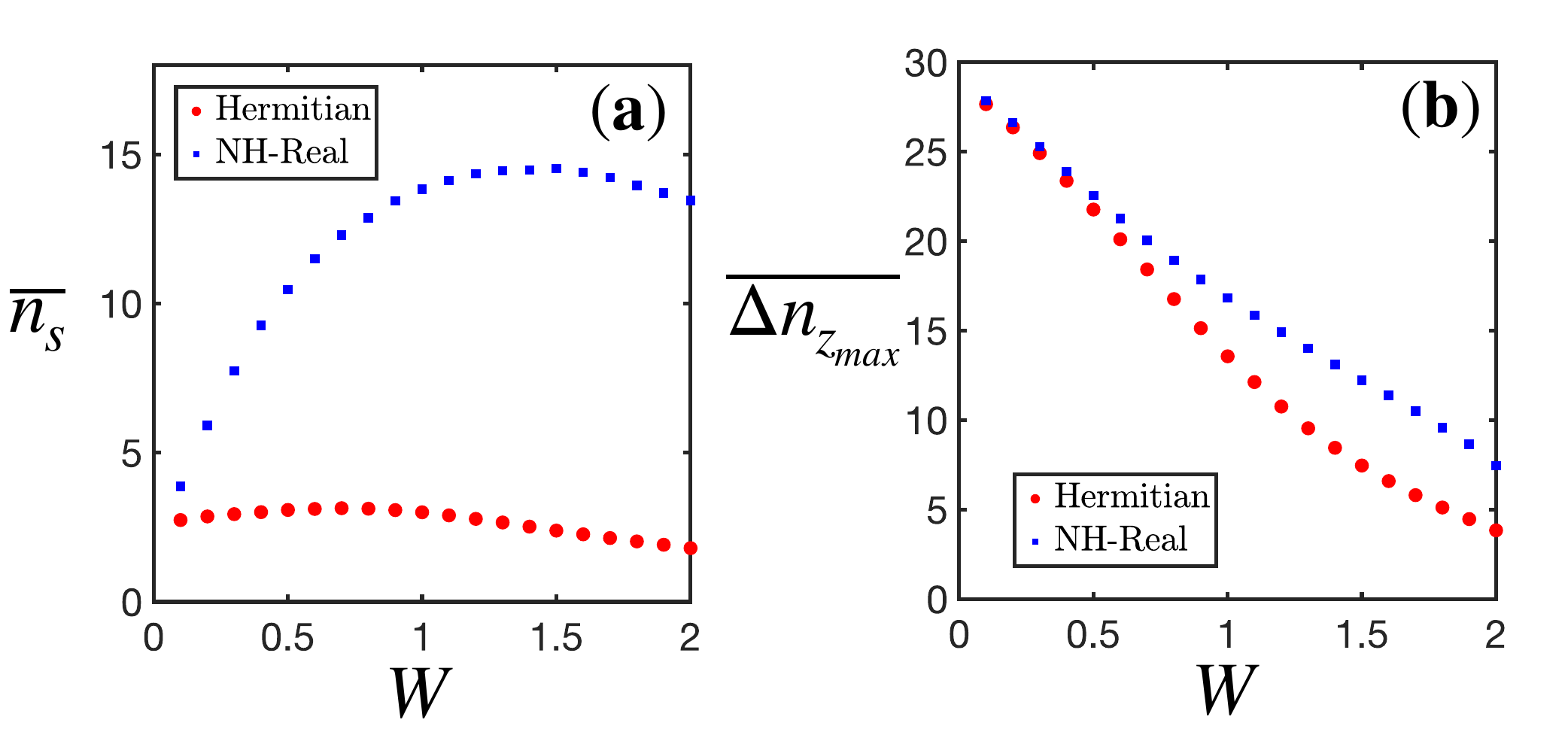}
    \caption{\textbf{Asymptotic study of propagation dynamics:} Dependence of (a) averaged absolute shift of the mean position $\overline{n_s}$ and (b) averaged uncertainty $\overline{\Delta n_{z_{\max}}}$ at $z_{\max}=2\cdot 10^4$ on the disorder strength $W$, for the \textit{Hermitian} case (red circles) and the \textit{real non-Hermitian} case (blue squares), for lattices of $N=100$ sites.}

    \label{fig_1}
\end{figure}

At this point it is important to comment on the evolution of the optical power $\mathcal{P}(z)$ in the real non-Hermitian off-diagonal disordered model. As noted earlier, this quantity is not conserved during propagation. However, in contrast to lattices whose Hamiltonian possesses a complex eigenspectrum, reflecting the presence of gain and loss, where the power typically grows exponentially due to the dominance of the most amplifying eigenstates, here the power variations are not associated with exponential increase. This behavior is illustrated in Fig.~5(a) and Fig.~5(b), which show the evolution of optical power for the realizations shown in Fig.~4(c) and Fig.~4(g), respectively (solid black lines), compared against their corresponding Hermitian cases of Fig.~4(a) and Fig.~4(e) (dotted lines).  
\begin{figure}[t]
    \centering
    \includegraphics[width=1\columnwidth]{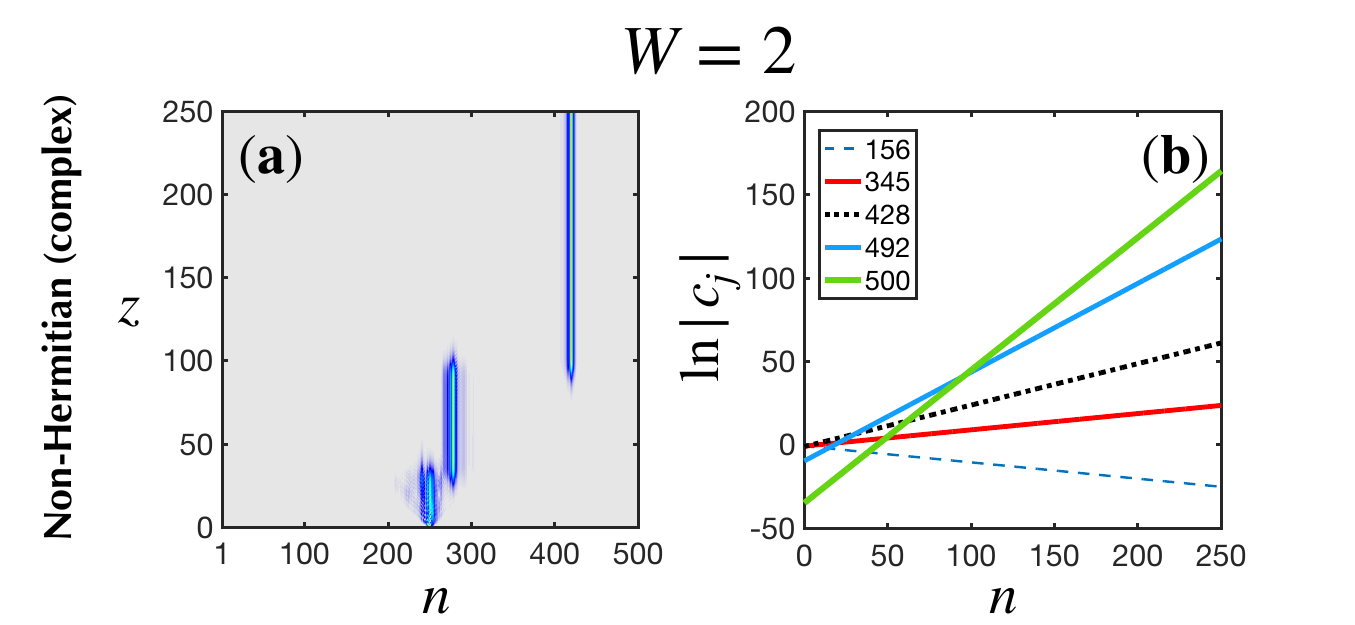}
    \caption{\textbf{Propagation dynamics for 1D lattices} ($N=500$ sites): (a) Evolution of normalized wavefunction initially localized at site $n=250$, for single strong disorder realizations ($W=2$), for the  \textit{complex non-Hermitian} case. (b) Evolution of the logarithmic magnitudes $\ln|c_j|$ of representative eigenmodes for the complex non-Hermitian case, sorted in descending order of the imaginary part of their eigenvalues (inset). }

    \label{fig_1}
\end{figure}
Remarkably, for the real non-Hermitian case we identify an invariant quantity, introduced in Appendix~A, which we denote as the \emph{pseudopower}:  
\begin{equation}
   \tilde{\mathcal{P}}(z) \;=\; |\psi_1(z)|^2 \;+\; \sum_{n=2}^{N} 
   \left( \prod_{m=1}^{n-1}\frac{t_{R,m}}{t_{L,m}} \right) |\psi_n(z)|^2 .
\end{equation}
In both panels of Fig.~5 the corresponding values of pseudopower are included, demonstrating its exact invariance. This conservation law provides a direct means of validating the accuracy of our numerical simulations, particularly given the strong non-normality inherent in non-Hermitian lattices with exclusively off-diagonal elements.

Up to this point we have only studied the propagation dynamics for single disorder realizations. To obtain a more general understanding of the transport differences between the Hermitian and real non-Hermitian off-diagonal disordered lattices, we perform a statistical analysis over 50000 realizations, focusing on two \emph{asymptotic} quantities for lattices with $N=100$ under $\psi_{n}=\delta_{n,50}$: the averaged \emph{absolute shift} of the mean position at $z_{max}\equiv 20000$, i.e. $\overline{n_{s}}\equiv \overline{|\langle n \rangle_{z_{max}}-\langle n \rangle_{0}|}$, as well as the averaged uncertainty for the same propagation distance, i.e. $\overline{\Delta n_{z_{max}}}$; we then compare the Hermitian and non-Hermitian cases as the disorder strength $W$ varies. As shown in Fig.~6(a), the absolute shift $\overline{n_{s}}$ is systematically larger in the non-Hermitian regime, reflecting asymmetric delocalization of the wavefunction away from the initial excitation region. This shift increases monotonically up to $W\approx 1.4$ and then exhibits a slight decrease. Although this decrease may appear counterintuitive (one might expect stronger disorder to promote jumps to distant regions) it can be attributed to the dual character of off-diagonal disorder: it simultaneously enhances localization while permitting highly preferential hopping paths that enable delocalization, leading to a nontrivial interplay. By contrast, for small $W$ the averaged asymptotic uncertainties are essentially identical in the two models; as $W$ increases, their difference grows, with larger values for the non-Hermitian case, indicating either enhanced spreading [as in Fig.~4(c)] or localization about distant lattice localization centers (as in Fig.~4(h)).

Regarding the complex non-Hermitian off-diagonal disorder case, where the eigenvalue spectrum is complex, the dynamics are dictated by the evolution of the projection coefficients \( |c_j(z)| = |c_{j,0}| e^{-\mathrm{Im}(\epsilon_j)z} \), as shown in Eq.~\eqref{exp}. As discussed in detail in Ref.~\cite{leventis_2022}, in systems with complex spectra and localized eigenmodes, a jump occurs whenever the projection \( |c_k(z)| \) of an eigenstate \( \ket{u_{k}^{R}} \) becomes greater than the projection \( |c_r(z)| \) of the previously dominant eigenstate \( \ket{u_{r}^{R}} \), under the condition \( \mathrm{Im}(\epsilon_k) < \mathrm{Im}(\epsilon_r) \). This behavior is further confirmed for the complex off-diagonal disordered model shown in Fig.~7, for a single disorder realization with \( W=2 \).

\section{TWO-DIMENSIONAL OFF-DIAGONAL DISORDERED LATTICES}
\subsection{Hermitian and non-Hermitian models}
Having examined in detail the spectral and dynamical features of 1D lattices with off-diagonal disorder, we now proceed to the corresponding 2D lattice case. In this section, we consider a square 2D lattice consisting of \(N\times N\) evanescently coupled waveguides (indexed by \((n_x, n_y)\in\{1,2,\ldots,N\}\times\{1,2,\ldots,N\}\)) with uniform on-site terms \(\alpha_{n_{x},n_{y}}\equiv 0\) and site-dependent nearest-neighbor couplings. Thus, the Hamiltonian of the system is described by
\begin{equation}
\label{off2D}
\scriptsize
\begin{aligned}
H = \sum_{n_x,n_y} \Big(
& t^{(x)}_{L;x,y}\ket{n_x,n_y}\bra{n_x+1,n_y}
+ t^{(x)}_{R;x,y}\ket{n_x+1,n_y}\bra{n_x,n_y} \\
&+ t^{(y)}_{D;x,y}\ket{n_x,n_y}\bra{n_x,n_y+1}
+ t^{(y)}_{U;x,y}\ket{n_x,n_y+1}\bra{n_x,n_y} \Big).
\end{aligned}
\normalsize
\end{equation}
Here, \(t^{(x)}_{L;x,y}\) and \(t^{(x)}_{R;x,y}\) denote the coupling amplitudes to the backward and forward directions along $n_x$, respectively, while \(t^{(y)}_{D;x,y}\) and \(t^{(y)}_{U;x,y}\) denote the coupling amplitudes to the downward and upward directions along $n_y$, respectively. We assume that the waveguides form a square array with open boundary conditions (OBC), i.e., \(\ket{0,n_y}=\ket{N_x+1,n_y}=\ket{n_x,0}=\ket{n_x,N_y+1}\equiv 0\). In what follows, we distinguish between three different cases depending on the specific values of the coupling coefficients \(t^{(x)}_{L/R;x,y}\) and \(t^{(y)}_{D/U;x,y}\), in direct analogy with the 1D case.
\begin{figure}[t]
    \centering
    \includegraphics[width=1\columnwidth]{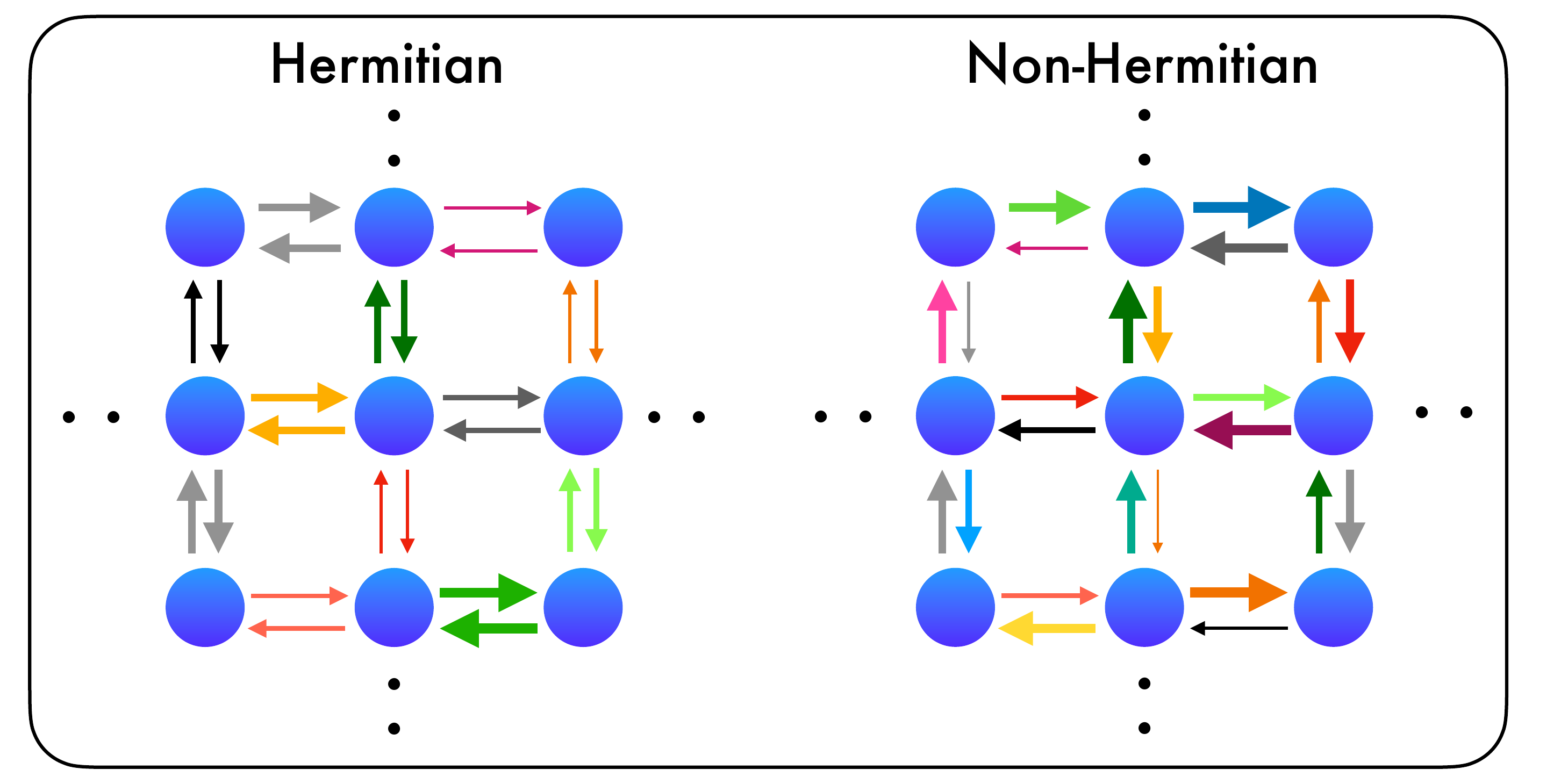}
    \caption{\textbf{Schematic representation} of the two-dimensional lattice models considered in this work.  (a) Random nearest-neighbor couplings are identical in the forward and backward directions, leading to a Hermitian Hamiltonian.  (b) Random nearest-neighbor couplings differ between forward and backward directions, giving rise to a non-Hermitian Hamiltonian. In both cases, variations in the color and width of the arrow indicate different coupling strengths, while in case (b) the couplings may be either real or complex.}

    \label{fig_1}
\end{figure}

In the \emph{Hermitian} off-diagonal disorder case, the couplings are identical in both directions on each bond,
\[
t^{(x)}_{L;x,y}=t^{(x)}_{R;x,y}\equiv t^{(x)}_{x,y},\qquad
t^{(y)}_{D;x,y}=t^{(y)}_{U;x,y}\equiv t^{(y)}_{x,y},
\]
where, as in the 1D case, \(t^{(x)}_{x,y}\) and \(t^{(y)}_{x,y}\) are random positive real numbers drawn from a uniform distribution \(t^{(\kappa)}_{x,y}\in\big[1-\tfrac{W}{2},\,1+\tfrac{W}{2}\big]\) (\(\kappa\in\{x,y\}\)). As shown in Appendix~B, the eigenvalue problem \(H\ket{u_j}=\epsilon_j \ket{u_j}\) also yields a \emph{chiral} spectrum \(\{\epsilon_j\}\), i.e., a real spectrum symmetric with respect to \(\epsilon=0\), with eigenmodes appearing in pairs and exhibiting identical amplitude distributions across the lattice. 

In analogy with the 1D case, for the non-Hermitian lattices we distinguish between two configurations. In the first, referred to as the \emph{real non-Hermitian} model, the couplings along opposite directions remain positive and real but are drawn independently from uniform distributions,
\[
t^{(x)}_{L;x,y},\,t^{(x)}_{R;x,y}\in\Big[1-\tfrac{W}{2},\,1+\tfrac{W}{2}\Big],
\]
\[
t^{(y)}_{D;x,y},\,t^{(y)}_{U;x,y}\in\Big[1-\tfrac{W}{2},\,1+\tfrac{W}{2}\Big].
\]
with \(W\in[0,2]\). The Hamiltonian is therefore real-valued but non-Hermitian due to the asymmetry of the couplings. Unlike its 1D analogue, however, in 2D the corresponding eigenvalue spectrum is \emph{complex}.

In the second configuration, which, as in the 1D scenario. we call the \emph{complex non-Hermitian} model, the couplings are not only unequal but also complex, defined as
\[
t^{(x)}_{L;x,y}=\frac{a^{(x)}_{x,y}+i\,b^{(x)}_{x,y}}{\sqrt{2}},\quad
t^{(x)}_{R;x,y}=\frac{c^{(x)}_{x,y}-i\,d^{(x)}_{x,y}}{\sqrt{2}},
\]
\[
t^{(y)}_{D;x,y}=\frac{a^{(y)}_{x,y}+i\,b^{(y)}_{x,y}}{\sqrt{2}},\quad
t^{(y)}_{U;x,y}=\frac{c^{(y)}_{x,y}-i\,d^{(y)}_{x,y}}{\sqrt{2}},
\]
where \(a^{(\kappa)}_{x,y}, b^{(\kappa)}_{x,y}, c^{(\kappa)}_{x,y}, d^{(\kappa)}_{x,y}\) (\(\mu\in\{x,y\}\)) are independent random real variables drawn from the uniform distribution \(\big[1-\tfrac{W}{2},\,1+\tfrac{W}{2}\big]\). 

For both non-Hermitian random-coupling cases, the spectrum is complex. Nevertheless, the chiral symmetry of the lattice persists: eigenvalues appear in symmetric pairs with respect to the origin in the complex plane, and the corresponding paired eigenstates exhibit identical amplitude distributions across the lattice sites. In addition, for the real non-Hermitian case, the reality of the Hamiltonian enforces an extra complex-conjugate pairing symmetry, such that all eigenvalues appear in quartets \(\{\epsilon_j,\, \epsilon_j^{*},\, -\epsilon_j,\, -\epsilon_j^{*}\}\). A schematic representation highlighting the differences between the Hermitian and non-Hermitian lattices is shown in Fig.~8. In the following subsections, we analyze and compare the spectral and transport properties of the finite-size (\(N\times N\)) 2D square off-diagonal disordered lattices introduced above, in analogy to the 1D case.

\subsection{Spectral properties}
\begin{figure*}[t]
    \centering
    \includegraphics[width=0.68\textwidth]{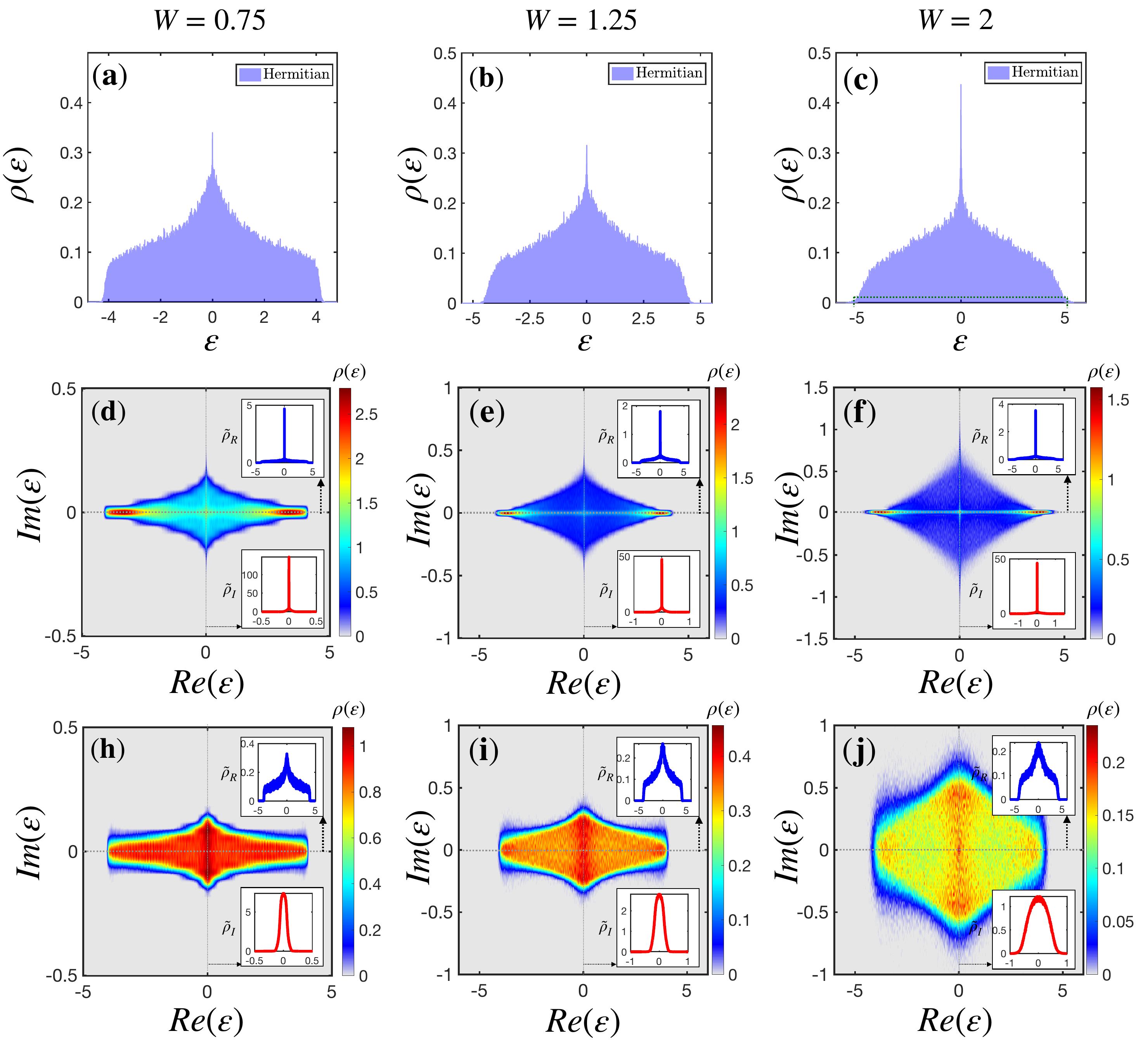}
    \caption{\textbf{Density of states for 2D lattices} ($N\times N=50\times 50$ sites): (a)–(c) DOS of real eigenvalues, $\rho(\epsilon)$, for the \textit{Hermitian} case (blue-shaded surface) with (a) $W=0.75$, (b) $W=1.25$, and (c) $W=2$. (d)–(f) / (h)–(j) DOS in the complex plane (colormap) for the \textit{real non-Hermitian} / \textit{complex non-Hermitian} case. \textit{Insets} show the real-projected DOS $\tilde{\rho}_R$ (blue lines) and the imaginary-projected DOS $\tilde{\rho}_I$ (red lines) for (d)/(h) $W=0.75$, (e)/(i) $W=1.25$, and (f)/(j) $W=2$.}
    \label{fig_1}
\end{figure*}
In this subsection, we examine the spectral properties of finite-sized 2D lattices with non-Hermitian off-diagonal disorder, with particular emphasis on the density of states along the real axis or complex plane, and the participation ratio of the associated eigenmodes. In the 2D case, the definition given in Eq.~\eqref{pr} can be directly generalized to
\begin{equation}
    P_{j} \equiv 
    \frac{\left( \sum_{n_x,n_y=1}^{N,N} \big| u^{R}_{j,(n_x,n_y)} \big|^2 \right)^{2}}
         {\sum_{n_x,n_y=1}^{N,N} \big| u^{R}_{j,(n_x,n_y)} \big|^{4}} ,
\end{equation}
where \(u^{R}_{j,(n_x,n_y)}\equiv \braket{n_x,n_y}{u^{R}_{j}}\), ranging from \(P_j=1\) for a state localized at a single site, i.e., \(u^{R}_{j,(n_x,n_y)}=\delta_{n_x,m_x}\delta_{n_y,m_y}\), up to \(P_j=N^2\) for a state uniformly distributed over the entire square lattice, i.e., \(u^{R}_{j,(n_x,n_y)}=1/N\).

First, in the top row of Fig.~9, we show for reference the DOS corresponding to the real spectrum of the 2D Hermitian lattice for various values of the disorder strength \(W\), using a lattice of \(N\times N=50\times 50\) waveguide channels. As evident in Fig.~9(a) for \(W=0.75\), the DOS exhibits a pronounced peak at \(\epsilon=0\) and decreases monotonically toward the unperturbed band edge \(|\epsilon|=4\). This central peak, which appears even for smaller disorder, is a clear fingerprint of the DOS of the unperturbed 2D tight-binding lattice. As the disorder strength increases [Figs.~9(b) and 9(c)], the DOS tails extend to higher eigenvalues while the central peak becomes increasingly pronounced. Regarding localization, the average participation ratios $\overline{P}$ are shown in the first row of Fig.~10, where, in contrast to the 1D Hermitian case, the states near \(\epsilon=0\) correspond to a local minimum, similar to what is observed in the non-Hermitian 1D cases. Specifically, the maximum values of $\overline{P}$ are obtained near \(\epsilon \approx \pm 1.7\), ranging from $\overline{P}\approx 700$ for \(W=0.75\) [Fig.~10(a)] to $\overline{P}\approx 500$ for \(W=1.25\) [Fig.~10(b)] and $\overline{P}\approx 200$ for \(W=2\) [Fig.~10(c)]. It should be noted that these values of $\overline{P}$, when compared to the total lattice size (\(N\times N=2500\)), reveal that 2D Hermitian lattices exhibit much weaker localization than their 1D counterparts (top row of Fig.~3).

The form of the DOS becomes considerably more intricate in the real non-Hermitian case, where, unlike the 1D scenario, the eigenvalues are complex. As shown in Fig.~9(d) for \(W=0.75\), the highest values of the DOS in the complex plane appear, rather unexpectedly, near the unperturbed band edges \(\epsilon=\pm 4\). A mid-band peak at \(\epsilon=0\) is also present, although many of the eigenvalues that contributed to this peak as \(W\to 0\) are now spread along the imaginary axis. This redistribution leads to a pronounced peak in the real-projected DOS \(\tilde{\rho}_R\) at \(\epsilon_R=0\). Nevertheless, the vast majority of eigenvalues remain close to the real axis, as indicated by the imaginary-projected DOS. As the disorder strength increases to \(W=1.25\) [Fig.~9(e)] and \(W=2\) [Fig.~9(f)], the main features of the DOS remain similar, but a growing fraction of eigenvalues spread deeper into the complex plane, acquiring larger imaginary parts, and \(\rho(\epsilon)\) broadens into a rhombus-like shape. Regarding localization, the average participation ratios $\overline{P}(\epsilon)$ are significantly larger than in the Hermitian case, indicating a stronger tendency toward delocalization. The largest values range from $\overline{P}\approx 1200$ for \(W=0.75\) [Fig.~10(d)] to $\overline{P}\approx 700$ for \(W=1.25\) [Fig.~10(e)] and $\overline{P}\approx 350$ for \(W=2\) [Fig.~10(f)], consistently occurring around \(\epsilon_R\approx \pm 1.7\). While dips near \(\epsilon=0\) are still present for \(W=0.75\) and \(W=1.25\), the \(W=2\) case is particularly striking: here the minimum reaches $\overline{P}\approx 50$, corresponding to the most localized states of the entire spectrum [Fig.~10(f)].

In contrast, the complex non-Hermitian model produces a DOS that much less strongly favors eigenvalues near the real axis and does not exhibit pronounced peaks or singularities. For \(W=0.75\) [Fig.~9(h)], the eigenvalues spread across an approximately rectangular region in the complex plane. The real-projected DOS \(\tilde{\rho}_R\) shows a mild peak near \(\epsilon_R=0\), but it is not sharply singular at \(\epsilon=0\), reflecting the distribution of eigenvalues along the imaginary axis. Meanwhile, the imaginary-projected DOS \(\tilde{\rho}_I\) is approximately Gaussian, extending over \(\epsilon_I\in(-0.1,0.1)\). For higher disorder [\(W=1.25\) in Fig.~9(i) and \(W=2\) in Fig.~9(j)], the imaginary parts broaden (roughly \(\epsilon_I\in(-0.3,0.3)\) and \(\epsilon_I\in(-0.8,0.8)\), respectively), and \(\tilde{\rho}_I\) develops a flat-topped Gaussian-like profile. Localization in the complex non-Hermitian case is even weaker than in the real non-Hermitian counterpart, as shown in the third row of Fig.~10, with systematically higher participation ratios that peak at around $\overline{P}\approx 950$ for \(W=1.25\) [Fig.~10(i)] and $\overline{P}\approx 670$ for \(W=2\) [Fig.~10(j)]. In all cases, a local dip of $\overline{P}$ near the origin persists, making it a rather general feature for non-Hermitian off-diagonal disordered lattices.

\begin{figure*}[t]
    \centering
    \includegraphics[width=0.68\textwidth]{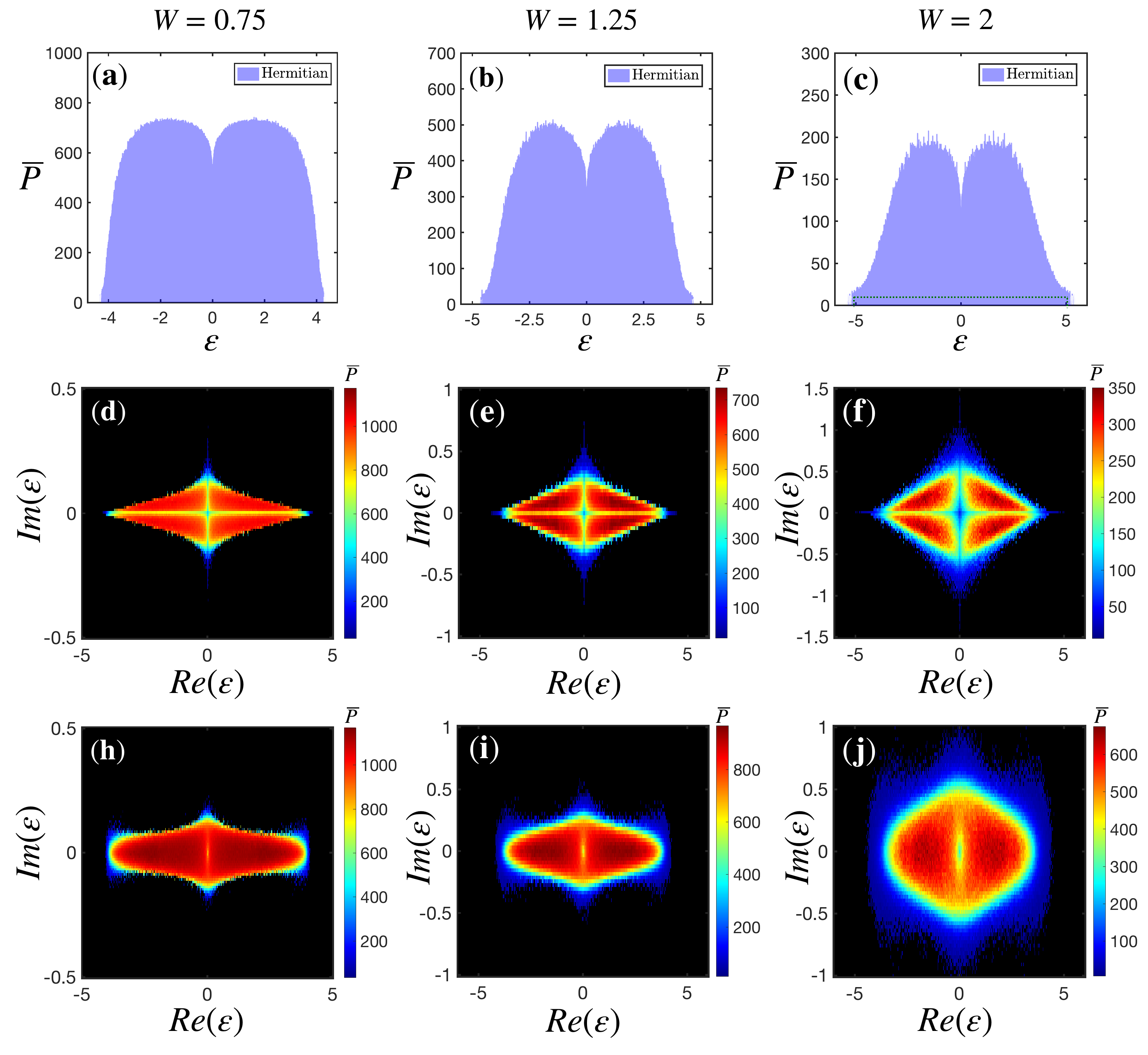}
    \caption{\textbf{Averaged participation ratios for 2D lattices} ($N\times N=50\times 50$ sites): (a)–(c) $\overline{P}(\epsilon)$ for eigenmodes in the \textit{Hermitian} case (blue-shaded surface) with (a) $W=0.75$, (b) $W=1.25$, and (c) $W=2$. (d)–(f) / (h)–(j) $\overline{P}(\epsilon)$ in the complex plane (colormap) for the \textit{real non-Hermitian} / \textit{complex non-Hermitian} case with (d)/(h) $W=0.75$, (e)/(i) $W=1.25$, and (f)/(j) $W=2$.}
    \label{fig_1}
\end{figure*}
\subsection{Propagation dynamics}
Having examined the spectral properties of the two-dimensional off-diagonal disordered lattices, we now turn to their transport features. The coupled-mode equation governing paraxial evolution in the lattice of Eq.~\eqref{off2D} reads
\begin{equation}
\normalsize
\begin{aligned}
i\,\frac{d \psi_{n_x,n_y}}{dz}
&+ t^{(x)}_{R;x,y}\,\psi_{n_x+1,n_y}
+ t^{(x)}_{L;x-1,y}\,\psi_{n_x-1,n_y} \\
&+ t^{(y)}_{U;x,y}\,\psi_{n_x,n_y+1}
+ t^{(y)}_{D;x,y-1}\,\psi_{n_x,n_y-1}
=0 ,
\end{aligned}
\normalsize
\end{equation}
where $\psi_{n_x,n_y}(z)$ is the amplitude of the electric field's envelope at site $(n_x,n_y)$. In the 2D case, the wavefunction is expressed as $\ket{\psi(z)}=\sum_{n_x=1}^{N}\sum_{n_y=1}^{N}\psi_{n_x,n_y}(z)\ket{n_x,n_y}$. In analogy with Eq.~\eqref{exp}, we expand the field in the right-eigenmode basis of $H$ as
\begin{equation}
    \label{exp_2d}
    \psi_{n_x, n_y}(z)=\sum_{j=1}^{N^2} c_{j,0}\, e^{\,i\epsilon_{j}z}\, u^{R}_{j,({n_x,n_y})},
\end{equation}
with coefficients $c_{j,0}=\braket{u_{j}^{L}}{\psi(0)}$. For non-Hermitian systems, we introduce a normalized amplitude for the field envelope at every propagation distance $z$, namely $\phi_{n_x,n_y}\equiv \psi_{n_x,n_y}/\sqrt{\mathcal{P}(z)}$, where $\mathcal{P}(z)\equiv\bra{\psi}\ket{\psi}=\sum_{n_x=1}^{N}\sum_{n_y=1}^{N}|\psi_{n_x,n_y}(z)|^2$.
\begin{figure*}[t]
    \centering
    \includegraphics[width=0.9\textwidth]{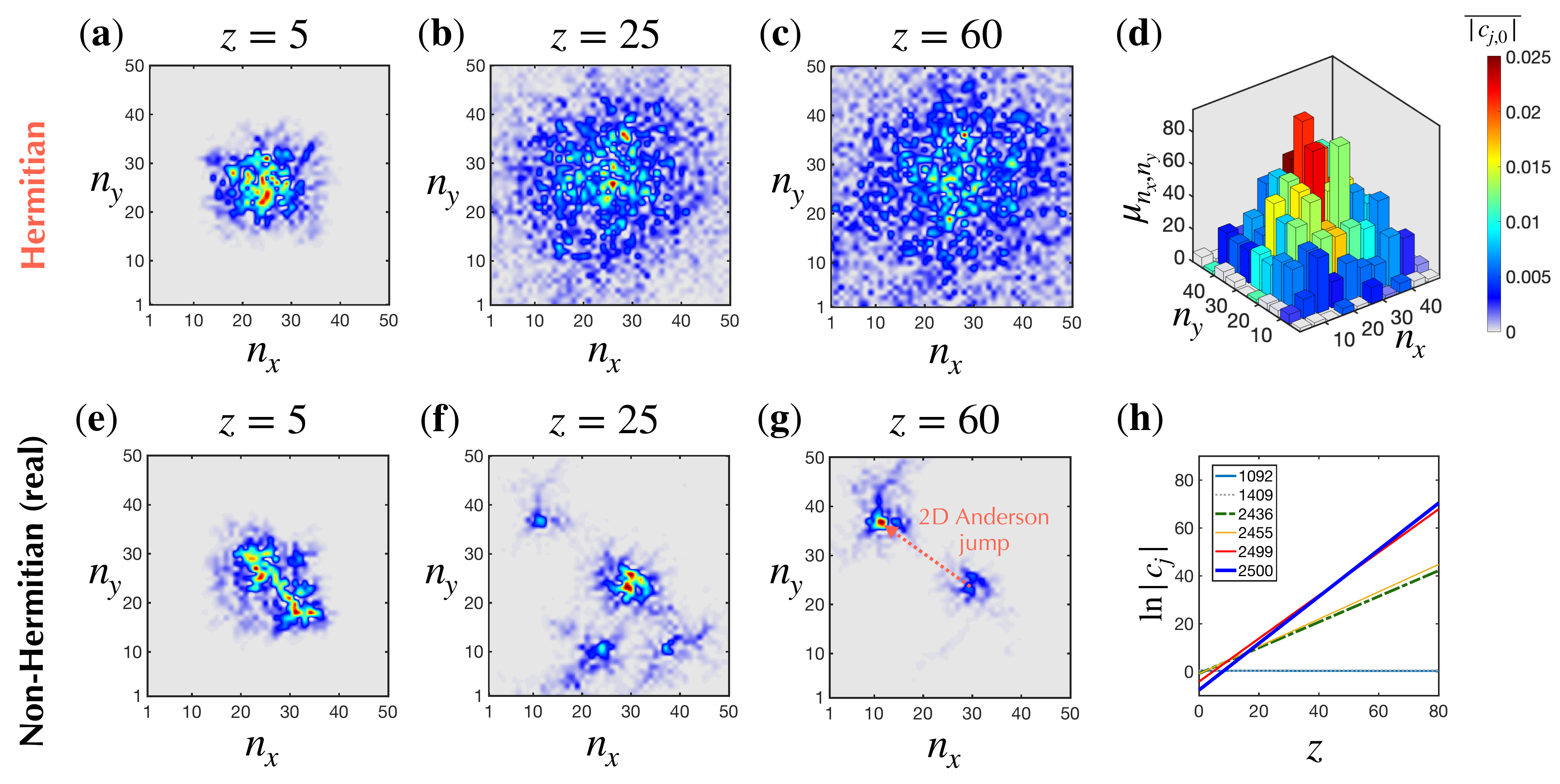}
    \caption{\textbf{Propagation dynamics for 2D lattices} ($N\times N=50\times 50$ sites): (a)–(c) / (e)–(g) Normalized wavefunction for single-site excitation at lattice site $(n_x,n_y)=(25,25)$ for single strong disorder realizations ($W=2$) in the \textit{Hermitian} / \textit{real non-Hermitian} cases, shown for (a)/(e) $z=5$, (b)/(f) $z=25$, and (c)/(g) $z=60$. Panel (d) shows the distribution $\mu_{n_x,n_y}$ of centers of mass of  eigenmodes for the Hermitian case (bar height), colored according to the averaged projection coefficients $\langle |c_{j,0}|\rangle$ of the eigenmodes within each bin, while panel (h) shows the evolution of the logarithmic magnitudes of representative eigenmodes for the non-Hermitian case, sorted in descending order of the imaginary part of their eigenvalues (shown in the inset).}
    \label{fig_1}
\end{figure*}

As discussed in detail in the 1D case, for lattices with real spectrum the evolution is dominated by modes $\ket{u_{j}^{R}}$ with large $|c_{j,0}|$, while eigenstates with negligible $|c_{j,0}|$ contribute only weakly to Eq.~\eqref{exp_2d}. In 2D lattices, for each right eigenstate $\ket{u_{j}^{R}}$, we can calculate its center of mass along the $n_x$ and $n_y$ directions,
\begin{equation}
    \langle m_{x/y} \rangle_{j} \equiv \frac{\sum_{n_{x/y}=1}^{N} n_{x/y}\,\abs{u_{j,(n_x,n_y)}^{R}}^2}{\sum_{n_{x/y}=1}^{N} \abs{u_{j,(n_x,n_y)}^{R}}^2}.
\end{equation}
In the top row of Fig.~11 we show the propagation dynamics in a 2D Hermitian off-diagonal disordered lattice ($W=2$) of $N\times N=50\times 50$ sites for different propagation distances [Figs.~10(a)–10(c)] following a single-channel excitation $\psi_{n_x,n_y}=\delta_{n_x,25}\delta_{n_y,25}$. Although the wavefunction spreads across the lattice, it never reaches a uniform distribution and instead remains peaked near the center. This is consistent with the spectral analysis of the Hamiltonian, which revealed weaker localization than in the 1D case. Moreover, this behavior reflects the fact that most eigenmodes have their centers of mass near the middle of the lattice, and, most importantly, that the central modes correspond to the largest projection coefficients $c_{j,0}$ [Fig.~11(d)].

Regarding the non-Hermitian cases, the dynamics differ dramatically from those of the conservative lattices. As discussed also for the 1D complex non-Hermitian case, due to the complex spectrum, the magnitudes of the projection coefficients are no longer constant but amplify or decay accoreding to \( |c_j(z)| = |c_{j,0}| e^{-\mathrm{Im}(\epsilon_j)z} \). Thus, their logarithms evolve as
\begin{equation}
    \ln |c_j(z)|=\ln |c_{j,0}|-\Im (\epsilon_j) z ,
\end{equation}
leading to \emph{amplification} for all eigenmodes with $\Im (\epsilon_j)<0$ and \emph{dissipation} for those with $\Im (\epsilon_j)>0$. Whenever the magnitude of a projection \( |c_k(z)| \) of an eigenstate \( \ket{u_{k}^{R}} \) exceeds the magnitude of the projection \( |c_r(z)| \) of the previously dominant eigenstate \( \ket{u_{r}^{R}} \), a \emph{two-dimensional Anderson jump} occurs. For a given disorder realization, the sequence of crossings between dominant eigenmodes, as well as the propagation distances at which they take place, depend on the initial projection coefficients $c_{j,0}$. Ultimately, however, the eigenstate associated with the largest gain [\( \min \{\mathrm{Im}(\epsilon_j)\} \)] governs the asymptotic dynamics. This unconventional transport behavior is illustrated in the bottom row of Figs.~11(e)–11(g) for a specific realization of real non-Hermitian off-diagonal disorder in a 2D lattice of $N\times N=50\times 50$ sites and $W=2$, where the normalized wavefunction is shown at different propagation distances. In Figure~11(h), we further depict the evolution of the logarithms of the projection coefficients, revealing the crossing points between successive dominant eigenmodes, which mark the occurrence of 2D Anderson jumps. Owing to the relatively weak localization of the eigenmodes in this model, whose tails spatially overlap, the states before and after a jump may share common regions. Finally, we note that the same analysis applies to the dynamics in 2D complex non-Hermitian off-diagonal disordered lattices, leading to the appearance of jumps for this lattice model as well.

\section{DISCUSSION AND CONCLUSIONS}
In this paper, we studied the spectral and dynamical features of one-dimensional and two-dimensional non-Hermitian off-diagonal finite-sized disordered optical lattices, i.e., lattices with randomness applied to the couplings rather than the on-site potential. More specifically, we analyzed the eigenvalue distribution in the complex plane and the degree of localization of the corresponding eigenmodes across different spectral regions, always in direct comparison with the properties of the respective Hermitian lattices. 

Furthermore, we examined and analyzed the propagation dynamics under single-channel excitation in such lattices, focusing on the asymptotic (i.e., long propagation distance) wavefunction. Our study revealed strong deviations between the non-Hermitian models and their conservative counterparts, such as non-Hermitian jumps due to amplification, which we first reported in 2D lattices, as well as non-Hermitian jumps in 1D lattices with nonreciprocal random hopping terms despite the fact the spectrum is \emph{purely real}.

We believe that our work provides a reference study on non-Hermitian off-diagonal disorder and may stimulate further research in non-Hermitian disordered physics, such as investigations of localization in the thermodynamic limit of the models introduced here using scaling or spectral criteria, including the possible existence of mobility edges or localization–delocalization transitions.

\textit{Note added}. After the submission of this manuscript, Ref. \cite{he2025} appeared, which discusses non-Hermitian jumps in disorder-free lattices with a purely real spectrum.
\section*{ACKNOWLEDGMENTS}
The authors acknowledge financial support from the European Research Council (ERC) through the Consolidator Grant Agreement No. 101045135 (Beyond Anderson).  

\appendix

\section{Spectral symmetries in 1D off-diagonal disordered lattices}
\label{app:1D}

In this Appendix we analyze the spectral symmetries underlying the one-dimensional (1D) off-diagonal disordered lattices introduced in Sec.~II~A. We consider successively the Hermitian, real non-Hermitian, and complex non-Hermitian configurations described by the Hamiltonian of Eq.~(1),
\begin{equation}
H=\sum_{n=1}^{N-1}\!\left(t_{L,n}\,|n\rangle\langle n+1|+t_{R,n}\,|n+1\rangle\langle n|\right),
\label{eq:app_H}
\end{equation}
with open boundary conditions (OBC), $|0\rangle=|N+1\rangle\equiv 0$.

\paragraph{Hermitian off-diagonal disorder.}
In the Hermitian configuration, the couplings are real and symmetric on each bond,
\begin{equation}
t_{L,n}=t_{R,n}\equiv t_n\in\mathbb{R}.
\end{equation}
The eigenvalue problem $H|\psi\rangle=\epsilon|\psi\rangle$ then reads in components as Eq.~(8) of the main text, with $t_{L,n}=t_{R,n}=t_n$. Because the matrix is tridiagonal with vanishing diagonal elements, it possesses a chiral symmetry generated by
\begin{equation}
\Gamma=\sum_{n=1}^{N}(-1)^n|n\rangle\langle n|, 
\qquad 
\Gamma^2=\mathbb{I},
\end{equation}
which anticommutes with $H$,
\begin{equation}
\{\Gamma,H\}=0.
\end{equation}
Therefore, if $H|\psi\rangle=\epsilon|\psi\rangle$, then
\begin{equation}
H(\Gamma|\psi\rangle)=-\epsilon(\Gamma|\psi\rangle).
\end{equation}
The eigenvalues thus appear in symmetric pairs $\{\epsilon_j,-\epsilon_j\}$ about $\epsilon=0$, and the paired states share identical amplitude distributions across the lattice,
\begin{equation}
|\psi_{j,n}|=|(\Gamma\psi_j)_n|, \qquad \forall n,
\end{equation}
so that the spectrum is purely real and chiral-symmetric.

\paragraph{Real non-Hermitian off-diagonal disorder.}
When the couplings remain real and positive but become directionally asymmetric,
$t_{L,n}\neq t_{R,n}$
the Hamiltonian of Eq.~\eqref{eq:app_H} is real but non-Hermitian, $H\neq H^\dagger$.

Under OBC, the spectrum remains entirely real. This can be shown by constructing a site-diagonal similarity transformation
\begin{equation}
S=\sum_{n=1}^{N}s_n|n\rangle\langle n|,
\qquad s_n>0,
\label{eq:S_def_1D}
\end{equation}
such that the transformed Hamiltonian
\begin{equation}
\tilde H = S^{-1} H S
\end{equation}
has symmetric nearest-neighbor couplings. Imposing symmetry on the transformed off-diagonal elements,
\begin{equation}
\tilde H_{n,n+1}=\tilde H_{n+1,n}\equiv \tau_n,
\end{equation}
gives the recursion
\begin{equation}
\tilde H_{n,n+1}
= s_n^{-1} t_{L,n} s_{n+1},
\qquad
\tilde H_{n+1,n}
= s_{n+1}^{-1} t_{R,n} s_n,
\end{equation}
and the symmetry condition $\tilde H_{n,n+1}=\tilde H_{n+1,n}$ implies
\begin{equation}
\frac{s_{n+1}}{s_n}
=\sqrt{\frac{t_{R,n}}{t_{L,n}}}.
\label{eq:1D_recursion}
\end{equation}
Under OBC there are no closed loops: starting from an arbitrary choice of $s_1>0$, Eq.~\eqref{eq:1D_recursion} uniquely and consistently determines all subsequent $s_n$,
\begin{equation}
s_n
= s_1 \prod_{m=1}^{n-1}\sqrt{\frac{t_{R,m}}{t_{L,m}}},
\qquad n=2,\dots,N.
\end{equation}
Hence the similarity transformation $S$ is always globally well-defined in 1D under OBC.

Substituting Eq.~\eqref{eq:1D_recursion} into the expressions for the transformed couplings yields
\begin{equation}
\tilde H_{n,n+1}
= s_n^{-1} t_{L,n} s_{n+1}
= \sqrt{t_{L,n} t_{R,n}}
\equiv \tau_n,
\end{equation}
and similarly
\begin{equation}
\tilde H_{n+1,n}
= s_{n+1}^{-1} t_{R,n} s_n
= \tau_n.
\end{equation}
Therefore $\tilde H$ is a Hermitian tridiagonal matrix with vanishing on-site terms and symmetric off-diagonal elements $\tau_n=\sqrt{t_{L,n}t_{R,n}}$. Since $H$ and $\tilde H$ are related by a similarity transformation, they are isospectral, and the eigenvalues of $H$ are all real:
\begin{equation}
\sigma(H)=\sigma(\tilde H)\subset\mathbb{R}.
\end{equation}
Moreover, $\tilde H$ anticommutes with the same chiral operator $\Gamma$, and thus
\begin{equation}
\{\Gamma,\tilde H\}=0
\quad\Rightarrow\quad
\{\Gamma,H\}=0,
\end{equation}
so the chiral symmetry and the $\epsilon\leftrightarrow -\epsilon$ pairing persist in the real non-Hermitian case. The corresponding chiral partner states again share identical amplitude distributions across the lattice.

Finally, the same similarity transformation $S$ that renders $\tilde H$ Hermitian also identifies a conserved quadratic form. Defining the transformed field
\begin{equation}
|\phi(z)\rangle = S^{-1}|\psi(z)\rangle,
\end{equation}
the evolution equation $-i\,\partial_z|\psi\rangle = H|\psi\rangle$ becomes
\begin{equation}
-i\,\partial_z|\phi\rangle
= S^{-1} H S |\phi\rangle
= \tilde H |\phi\rangle,
\end{equation}
so $\phi$ evolves under the Hermitian Hamiltonian $\tilde H$. The standard norm in this Hermitian frame,
\begin{equation}
\tilde{\mathcal{P}} \equiv \sum_{n=1}^{N} |\phi_n(z)|^2,
\end{equation}
is therefore conserved. In components, $\phi_n = \psi_n/s_n$, so
\begin{equation}
\tilde{\mathcal{P}}
= \sum_{n=1}^{N} \frac{|\psi_n|^2}{s_n^2}.
\end{equation}
From Eq.~\eqref{eq:1D_recursion} we have $s_{n+1}/s_n = \sqrt{t_{R,n}/t_{L,n}}$, which implies
\begin{equation}
\frac{1}{s_{n+1}^2}
= \frac{t_{R,n}}{t_{L,n}}\,\frac{1}{s_n^2}.
\end{equation}
Choosing $s_1=1$ without loss of generality (any constant prefactor would only rescale the invariant), this recursion yields
\begin{equation}
\frac{1}{s_n^2}
= \prod_{m=1}^{n-1}\frac{t_{R,m}}{t_{L,m}},
\qquad n\ge 2.
\end{equation}
Hence
\begin{equation}
\tilde{\mathcal{P}}
= |\psi_1(z)|^2
+\sum_{n=2}^{N}\left(\prod_{m=1}^{n-1}\frac{t_{R,m}}{t_{L,m}}\right)|\psi_n(z)|^2
\end{equation}
and since $\tilde{\mathcal{P}}$ is conserved under the evolution generated by $\tilde H$, the pseudopower $\tilde{\mathcal{P}}(z)$ is a conserved quantity for the original real non-Hermitian dynamics.

\paragraph{Complex non-Hermitian off-diagonal disorder.}
For complex asymmetric couplings,
\begin{equation}
t_{L,n}=\frac{a_n+i b_n}{\sqrt{2}}, \qquad 
t_{R,n}=\frac{c_n-i d_n}{\sqrt{2}},
\end{equation}
with $a_n,b_n,c_n,d_n$ independent real random variables uniformly distributed in $[1-W/2,1+W/2]$, the Hamiltonian becomes genuinely non-Hermitian and its spectrum generally complex. Nevertheless, Eq.~\eqref{eq:app_H} still anticommutes with $\Gamma$, guaranteeing that all eigenvalues occur in pointwise symmetric pairs about the origin of the complex plane,
\begin{equation}
\epsilon_j \leftrightarrow -\epsilon_j.
\end{equation}
The paired states again share identical site amplitudes,
\begin{equation}
|\psi_{j,n}|=|(\Gamma\psi_j)_n|,
\end{equation}
so chiral partners have identical spatial intensity profiles even though the spectrum is complex.

\section{Spectral symmetries in 2D off-diagonal disordered lattices}
\label{app:2D}

We now examine the spectral symmetries of the two-dimensional (2D) off-diagonal disordered lattices discussed in Sec.~III~A. The Hamiltonian under open boundary conditions is given by
\begin{equation}
\label{eq:2dH}
\scriptsize
\begin{aligned}
H = \sum_{n_x,n_y} \Big(
& t^{(x)}_{L;n_x,n_y}\ket{n_x,n_y}\bra{n_x+1,n_y}
+ t^{(x)}_{R;n_x,n_y}\ket{n_x+1,n_y}\bra{n_x,n_y} \\
&+ t^{(y)}_{D;n_x,n_y}\ket{n_x,n_y}\bra{n_x,n_y+1}
+ t^{(y)}_{U;n_x,n_y}\ket{n_x,n_y+1}\bra{n_x,n_y} \Big),
\end{aligned}
\normalsize
\end{equation}
where the nearest-neighbor couplings along $x$ and $y$ are random and may be real or complex. As in 1D, we distinguish the Hermitian case, the real non-Hermitian case, and the complex non-Hermitian case.

The 2D lattice admits a chiral operator defined as
\begin{equation}
\Gamma = \sum_{n_x,n_y} (-1)^{n_x+n_y} \ket{n_x,n_y}\bra{n_x,n_y},
\qquad \Gamma^2 = \mathbb{I}.
\end{equation}
Because each coupling in Eq.~\eqref{eq:2dH} connects sites whose coordinates differ by one in either direction, the operation of $H$ always flips the sign of $(-1)^{n_x+n_y}$. As a result, the Hamiltonian anticommutes with $\Gamma$,
\begin{equation}
\{\Gamma,H\}=0.
\label{eq:chiral2d}
\end{equation}
This relation establishes the chiral (sublattice) symmetry of the system. Consequently, if
\begin{equation}
H|u^R\rangle=\epsilon |u^R\rangle,
\end{equation}
then
\begin{equation}
H(\Gamma|u^R\rangle)=-\epsilon(\Gamma|u^R\rangle),
\end{equation}
implying that eigenvalues appear in symmetric pairs $\epsilon \leftrightarrow -\epsilon$. Thus, eigenvalues occur in symmetric pairs $\epsilon\leftrightarrow -\epsilon$ about the origin in the complex plane for all three cases—Hermitian, real non-Hermitian, and complex non-Hermitian.

Because $\Gamma$ is diagonal with entries $\pm 1$, it follows that
\begin{equation}
\big|(\Gamma u^R)_{n_x,n_y}\big| = \big|u^R_{n_x,n_y}\big| \quad \forall (n_x,n_y),
\label{eq:amplitudes}
\end{equation}
showing that chiral partners share \emph{identical amplitude distributions} across all lattice sites.

For the real non-Hermitian case, when all couplings are real, the Hamiltonian satisfies $H^* = H$. If
\begin{equation}
H|u^R\rangle = \epsilon |u^R\rangle,
\end{equation}
then complex conjugation gives
\begin{equation}
H \ket{u^R}^{*} = \epsilon^* \ket{u^R}^{*}.
\end{equation}
Combining this property with the chiral relation in Eq.~\eqref{eq:chiral2d} yields the \emph{quartet symmetry}
\begin{equation}
\{\epsilon,\, \epsilon^*,\, -\epsilon,\, -\epsilon^*\},
\label{eq:quartet}
\end{equation}
with associated right-eigenvectors 
\begin{equation}
\{\,\ket{u^R},\, \ket{u^R}^{*},\, \Gamma \ket{u^R},\, \Gamma \ket{u^R}^{*}\}.
\end{equation}
Each member of this quartet exhibits the same spatial amplitude pattern, since
\begin{equation}
|\Gamma u^{R}_{n_x,n_y}|=|u^{R}_{n_x,n_y}|,
\qquad
|u^{R*}_{n_x,n_y}|=|u^{R}_{n_x,n_y}|.
\end{equation}

In one dimension, a diagonal similarity transformation can always be constructed to symmetrize the hopping amplitudes and render the spectrum real under OBC, as shown in Appendix~\ref{app:1D}. In 2D, one may attempt a site-diagonal transformation
\begin{equation}
S = \sum_{n_x,n_y} s_{n_x,n_y} \ket{n_x,n_y}\bra{n_x,n_y},
\end{equation}
with
\begin{equation}
\frac{s_{n_x{+}1,n_y}}{s_{n_x,n_y}} = 
\sqrt{\frac{t^{(x)}_{R;n_x,n_y}}{t^{(x)}_{L;n_x,n_y}}}, 
\qquad
\frac{s_{n_x,n_y{+}1}}{s_{n_x,n_y}} = 
\sqrt{\frac{t^{(y)}_{U;n_x,n_y}}{t^{(y)}_{D;n_x,n_y}}}.
\label{eq:loop}
\end{equation}
However, consistency around each plaquette requires that
\begin{equation}
\frac{t^{(x)}_{R;n_x,n_y}\,t^{(y)}_{U;n_x{+}1,n_y}}
     {t^{(x)}_{L;n_x,n_y}\,t^{(y)}_{D;n_x,n_y{+}1}} = 1
\quad \text{for all } (n_x,n_y).
\label{eq:plaquette}
\end{equation}
This constraint is in general violated since the couplings are drawn independently, implying that no diagonal similarity transformation $S$ can globally remove the random asymmetry. Consequently, unlike in one dimension, the spectrum of the 2D real non-Hermitian model is generically complex under OBC.

For the complex non-Hermitian case, the reality condition $H^* = H$ no longer holds, and the conjugate-pairing symmetry is lost. Nevertheless, the chiral symmetry of Eq.~\eqref{eq:chiral2d} remains valid. Therefore, even though the spectrum is complex, eigenvalues still appear in symmetric pairs $\epsilon \leftrightarrow -\epsilon$, and their corresponding eigenstates share identical amplitude profiles as in Eq.~\eqref{eq:amplitudes}.

\end{document}